\documentclass[article]{aa}
\usepackage{graphicx}% Include figure files

\usepackage{bm}% bold math
\usepackage{natbib}
\usepackage{txfonts}
\usepackage{url}
\newcommand{\CaII }{Ca\,II~}
\begin{document}
\bibliographystyle{aa}
\input epsf

\title{The solar chromosphere at high resolution with IBIS}
\subtitle{ I. New insights from the \CaII 854.2 nm line}

\author{G. Cauzzi\inst{1}
\and K. P. Reardon\inst{1}
\and H. Uitenbroek\inst{2}
\and F. Cavallini\inst{1}
\and A. Falchi\inst{1}
\and R. Falciani\inst{3}
\and K. Janssen\inst{1}
\and T. Rimmele\inst{2}
\and A. Vecchio\inst{1}
\and F. W\"oger\inst{2,4}}

\institute{INAF - Osservatorio Astrofisico di Arcetri, I-50125 Firenze, Italy
\and National Solar Observatory, P.O. Box 62, Sunspot NM 88349, USA
\and Dipartimento di Astronomia, Universit\`a di Firenze, I-50125 Firenze, Italy
\and Kipenheuer Institute f\"ur Sonnenphysics,  D-79104 Freiburg, Germany}
%\email{gcauzzi@arcetri.astro.it}
\date{\today}

\abstract{The chromosphere remains a poorly understood part of the solar atmosphere, as 
current modeling and observing capabilities are still ill-suited to investigate in depth its fully 
3-dimensional nature. In particular, chromospheric observations that can preserve high spatial 
and temporal resolution while providing spectral information over extended fields of view are still 
very scarce.}
{In this paper, we seek to establish the suitability of imaging spectroscopy performed 
in the \CaII 854.2 nm  line as a means to investigate the solar chromosphere at high resolution.}
{We 
utilize monochromatic images obtained with the Interferometric BIdimensional Spectrometer 
(IBIS) at multiple wavelengths within the \CaII 854.2 nm line and over several quiet areas. We analyze both the morphological properties derived from 
narrow-band monochromatic images and the average spectral properties of distinct solar features 
such as network points, internetwork areas and fibrils.}
{The spectral properties derived over 
quiet-Sun targets are in full agreement with earlier results obtained with fixed-slit spectrographic 
observations, highlighting  the reliability of the spectral information obtained with IBIS. Furthermore, the 
very narrowband IBIS imaging  reveals with much clarity the dual nature of the \CaII 854.2 nm line: its 
outer wings gradually sample the solar photosphere, while the core is a purely chromospheric indicator. 
The latter displays a wealth of fine structures including bright points, akin to the \CaII H$_{2V}$ and 
K$_{2V}$ grains, as well as fibrils originating from even the smallest magnetic elements. The fibrils occupy a large 
fraction of the observed field of view even in the quiet regions, and clearly outline 
atmospheric volumes with different dynamical properties, strongly dependent on the local magnetic 
topology. This highlights the fact that 1-D models stratified along the vertical direction can provide only a very 
limited representation of the actual chromospheric physics. }
{Imaging spectroscopy in the \CaII 854.2 nm line
currently represents one of the best observational tools to investigate the highly structured and highly dynamical 
chromospheric environment. A high performance instrument such as IBIS is crucial in order to achieve 
the necessary spectral purity and stability, spatial resolution, and temporal cadence. }
\maketitle
\titlerunning{The solar chromosphere in \CaII 854.2 nm}
\authorrunning{Cauzzi et al.}

\section{Introduction}

The chromosphere embodies the transition  between the photosphere and the corona, two regions  
dominated by vastly different physical regimes.    
In particular, it is within the chromosphere that the plasma $\beta$, the ratio of plasma kinetic pressure to 
magnetic pressure, falls below unity, signaling a shift from hydrodynamic to magnetic forces as the 
dominant agent in the structuring of the atmosphere. As ably described in the review of \citet
{2006ASPC..354..259J},
% Phil Judge review, Sac Peak 2005
the combined effects of magnetic field guidance and small scale gas thermodynamics  lead to the 
impressive amount of fine structures (jets, spicules, fibrils, mottles, etc.) that uniquely characterize this 
part of the solar atmosphere. Such fine structure represents a formidable challenge even to the most 
modern instrumentation, as its study requires high
spectral resolution, necessary to resolve line profiles encoding large gradients or discontinuities \citep
[e.g. the shocks of][]{1997ApJ...481..500C},
% Carlsson & Stein, shocks in K2v
 combined with extremely high temporal and spatial resolution 
\citep[see e.g. the recent observations by][]{2006ApJ...648L..67V}.
%van Noort & Rouppe 2006, ApJ letter

Among the chromospheric diagnostics accessible to ground-based observations, the
\CaII H and K resonance lines have been used most extensively
\citep[see e.g. the review of][]{1991SoPh..134...15R}.
%  Rutten \& Uitenbroek Sol Phys 1991. 
These lines are the broadest lines in the visible spectrum,
sampling a large range in formation height, and are the only
visible lines that provide a direct indication of the chromospheric
temperature rise with their H$_2$ and K$_2$ emission reversals.
Because the lines originate from the ground state of a dominant
ionization stage, they are mostly collisionally controlled in the
lower chromosphere, and sensitive to local temperature
to a much larger degree than for instance the hydrogen Balmer lines.
Of these, the  H$\alpha$ line in particular has also been widely exploited,
especially following the development of the Lyot filter \citep{lyot_33}, which gave rise to the science of `solar cinematography', i.e. narrow-band imaging obtained at rapid cadence.
Even though H$\alpha$ spectra are difficult to interpret because of
the complicated formation characteristics of this line,
much of what we know about the chromosphere
derives from the detailed H$\alpha$ morphology observed through
Lyot-style filters.

A somewhat neglected, but no less interesting, chromospheric diagnostics is  represented by the \CaII  
infrared triplet ($\lambda$ = 849.8, 854.2, 866.2 nm, hereafter \CaII IRT).  These lines originate
from transitions between the upper 4p $^2$P$_{1/2,3/2}$ levels and the lower metastable 3d $^2$D$_{3/2,5/2}$ levels. Transitions between the ground state 4s $^2$S$_{1/2}$ of \CaII and the  same upper 
levels give rise to the H and K lines. 
Since the branching ratio, the ratio between the spontaneous emission
coefficients of the H and K lines and those of the IRT lines, is about
$A_{\mathrm{HK}}/A_{\mathrm{IRT}} \equiv 15$, most photons emitted in
the IRT result from excitation in the H and K lines, leading to a very similar
temperature sensitivity for both sets of lines.
In addition, the 3d $^2$D$_{3/2,5/2}$ are metastable (i.e., there are
no allowed electric dipole transitions with the 4s $^2$S$_{1/2}$ ground state),
so that these levels can only be populated from below by collisional
excitation, strengthening the sensitivity of the IRT to local temperature
even more.

 First detected in the 1878 eclipse spectrum  by C. Young \citep[][]{1973SoPh...29...23E}, 
 %Eddy, 1973 research note
 the \CaII IRT has remained essentially ignored in solar observational work for nearly a century. From the 
1960's and up through the mid-1980's  it was occasionally employed, in particular in multi-line spectrographic 
studies \citep[among 
 others,][]{1968SoPh....5..131P,1970SoPh...11..374L,1971SoPh...20....3M,%
1972SoPh...25..357S,1972SoPh...25...81B,1974SoPh...39...49S,1984ApJ...277..874L}.
% Pasachoff et al 1968;  Mein 1971; Lites 1984; Shine linsky et al, various papers,
%% Beckers et al HIRKHAD program.
Spurred by the availability of CCD detectors with high sensitivity in
this wavelength range, the last decade has instead seen a rapid
 growth of observational studies adopting these lines both for solar
 and cool star research; we refer the reader to
 \citet{2006SoPh..235...55S,2006ApJ...639..516U,2006A&A...456..689T,%
2006ApJ...645..776U,2007ApJ...663.1386P}
 %Socas Navarro 2006 - SPINOR; Uitenbroek 2006 inverse C shape;
 % Tziotziou MSDP spot oscillation, Uitenbroek et al. 2006 siphon, Anna's thesis second paper
 for recent examples of solar studies, and to \citet{2000A&A...353..666C,2005A&A...430..669A}
%Chmielewski, 2000;  Andretta et al 2005 
for the stellar case. 

Despite the fact that near infrared wavelengths afford both a higher photon flux and a reduced terrestrial 
atmospheric disturbance (with respect to the wavelengths of H and K lines), high resolution solar
observations of the \CaII IRT are still scarce.  The great majority of the previous observations were in fact 
conducted with fixed-slit spectrographs, either performing an area raster scan,
requiring long repetition times and thus compromising the temporal resolution  
\citep{1994chdy.conf..103F,2006ApJ...639..516U,2007ApJ...663.1386P}, 
%Fleck Oslo 94, Uitenbroek 2006 inverse C shape, Anna's thesis second paper
 or keeping the slit at a fixed position  
 \citep{1990A&A...228..506D,2007ApJ...663.1386P,langangen_07}.
 % Deubner & Fleck 1990, cell-boundary distinction; Anna's thesis second paper; Oysten submitted
The latter strategy allows for a high temporal cadence, but makes it difficult to precisely follow the 
small chromospheric structures, often of magnetic origin, that are moved away from the slit either by 
atmospheric turbulence or solar evolution. A more efficient observational strategy is provided by  the 
Multichannel Subtractive Double Pass technique \citep[MSDP,][]{1991A&A...248..669M,2002A&A...381..271M}.
% Mein 1991, MSDP@VTT; Mein 2002, MDSP@THEMIS
The  MSDP is in fact one of the few examples of an Integral Field Unit (IFU) device being used for optical 
solar observations, 
with the unique capability of acquiring  truly simultaneous spectra over an extended field
of view. However, for the case of the  \CaII 854.2 nm line, observations are usually limited to a spectral 
range of about $\pm$ 40 pm from line core (e.g. Tziotziou et al 2002), thus possibly neglecting important 
information related to strong variations  encoded in the extended line profile. Moreover,  the usable field 
of view (FOV) represents a trade-off with the spectral coverage and resolution, and is generally limited to 
only a few arcsec in one of the spatial directions. 

In this framework, we introduce here new high resolution observations of the  \CaII
854.2 nm line obtained in a variety of solar structures with the Interferometric
BIdimensional Spectrometer \citep[IBIS,][]{2006SoPh..236..415C},
%% IBIS paper, Cavallini 2006
installed at the Dunn
Solar Telescope of the US National Solar Observatory. While not an IFU proper,
IBIS is a high performance instrument that combines most of the advantages of a full spectroscopic
analysis, usually obtained with single-slit spectrographs, with the high spatial
resolution, high temporal cadence and large field of view typical of filter
instruments.  Such characteristics are necessary to obtain new insights into
the structure and dynamics of the chromosphere \citep[see for example][]{2007A&A...461L...1V,2007ASPC..368..127C}.
%Vecchio et al 2007, letter on shadows
 %Cauzzi Coimbra. 
Although other instruments similar to IBIS are currently operative, such as TESOS  \citep{2002SoPh..211...17T}
%TESOS, upgrade to triple
and the new G\"ottingen Fabry-Perot system \citep{2006A&A...451.1151P}, 
%Puschmann et al 2006
no other imaging spectrometer is presently able to access the wavelengths of the \CaII IRT.

\section{Instrumental characteristics} \label{s_inst}

\begin{figure}[ht]
\includegraphics[width=8.4cm,height=13.5cm]{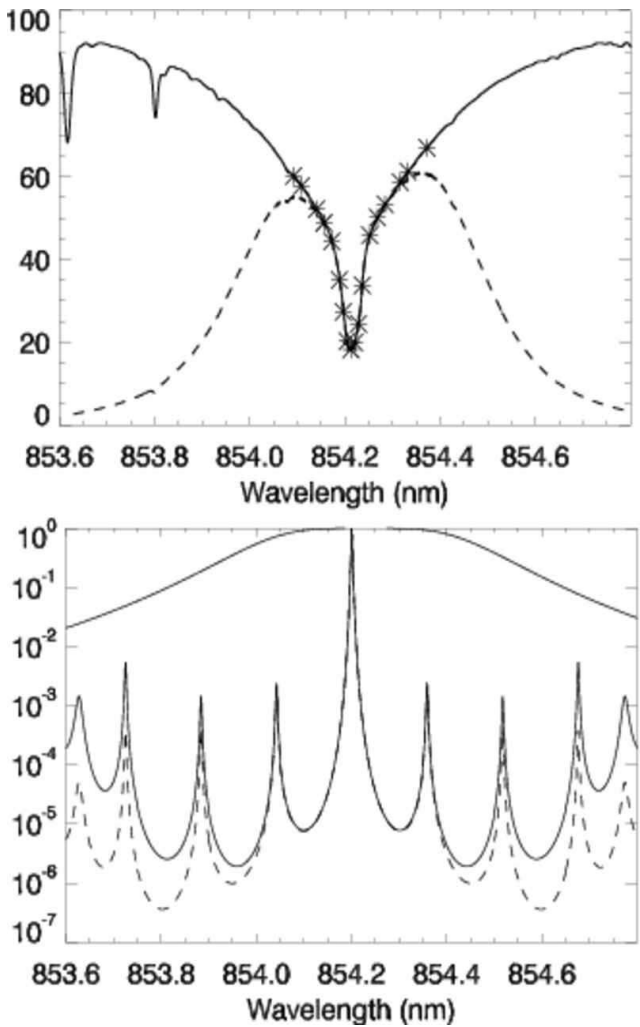}%{figpref.2.eps}
\caption{Top panel: the solid line displays the \CaII 854.2 nm atlas profile, while the dashed line shows the effects of the 0.46 nm FWHM prefilter utilized in IBIS observations. The actual maximum transmission of the prefilter is 35\%.  The asterisks indicate the wavelengths of  the images in Fig. \ref{f_20040531} and \ref{f_20040602}. Bottom panel: normalized prefilter profile (solid, upper curve) and periodic transmission profile of the combined Fabry Perot's (solid, multiple peaks curve). Note the logarithmic scale. The dashed line indicates the effects of the prefilter on the FPs transmission profile, i.e. the large suppression of the secondary maxima of the curve, necessary to reduce the spectral parasitic light.}
\label{f_inst_profile}
\end{figure}

\begin{figure*}
\includegraphics[width=18cm,height=15.75cm]{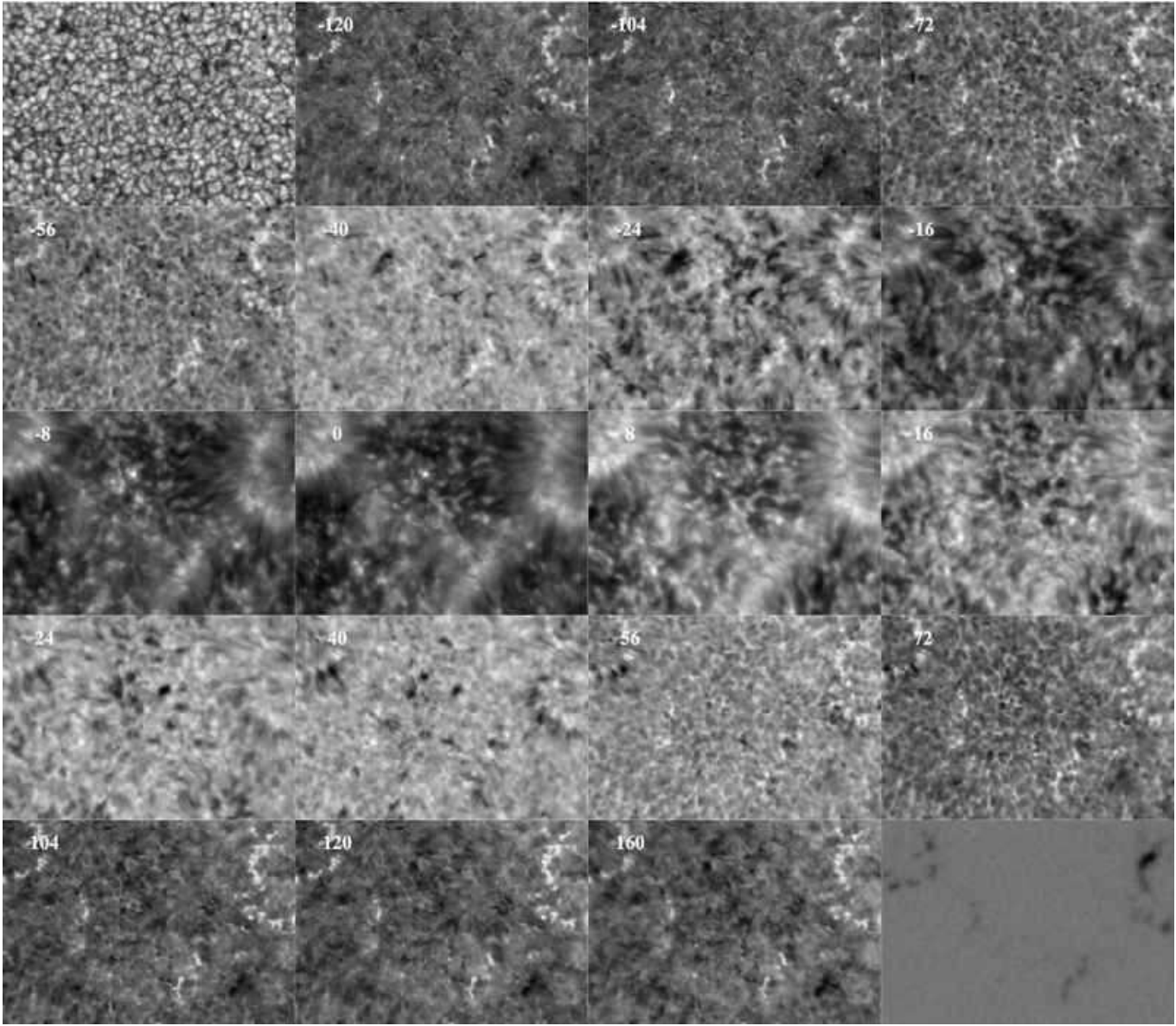}%{20040531.eps}
\caption{\CaII 854.2 nm data acquired on May 31, 2004. Wavelength runs from left to
right, top to bottom, as indicated in the panels, with offsets in pm from nominal line core. 
The figure shows a cutout of 64$''\times$41$''$
%380 x 240 pixels
of the full circular FOV, originally 80$"$ in diameter.
Each panel was
normalized for optimum display. %Tick marks in arcsec.
The topmost left panel shows the cospatial, cotemporal broadband image (acquired
at 710 nm, FWHM=10 nm). The bottom right panel gives the cotemporal HR MDI map,
scaled between $\pm$500 G.
The intensity in the far wings of the \CaII 854.2 nm line 
is strongly sensitive to 
the presence of magnetic structures \citep[cf.][and Sect. \ref{s_wings_mag}]{2006A&A...452L..15L}}.
%Leenaarts et al 2006, synthesis in MHD simulation 
\label{f_20040531}
\end{figure*} 

\begin{figure*}
\includegraphics[width=18cm,height=15.75cm]{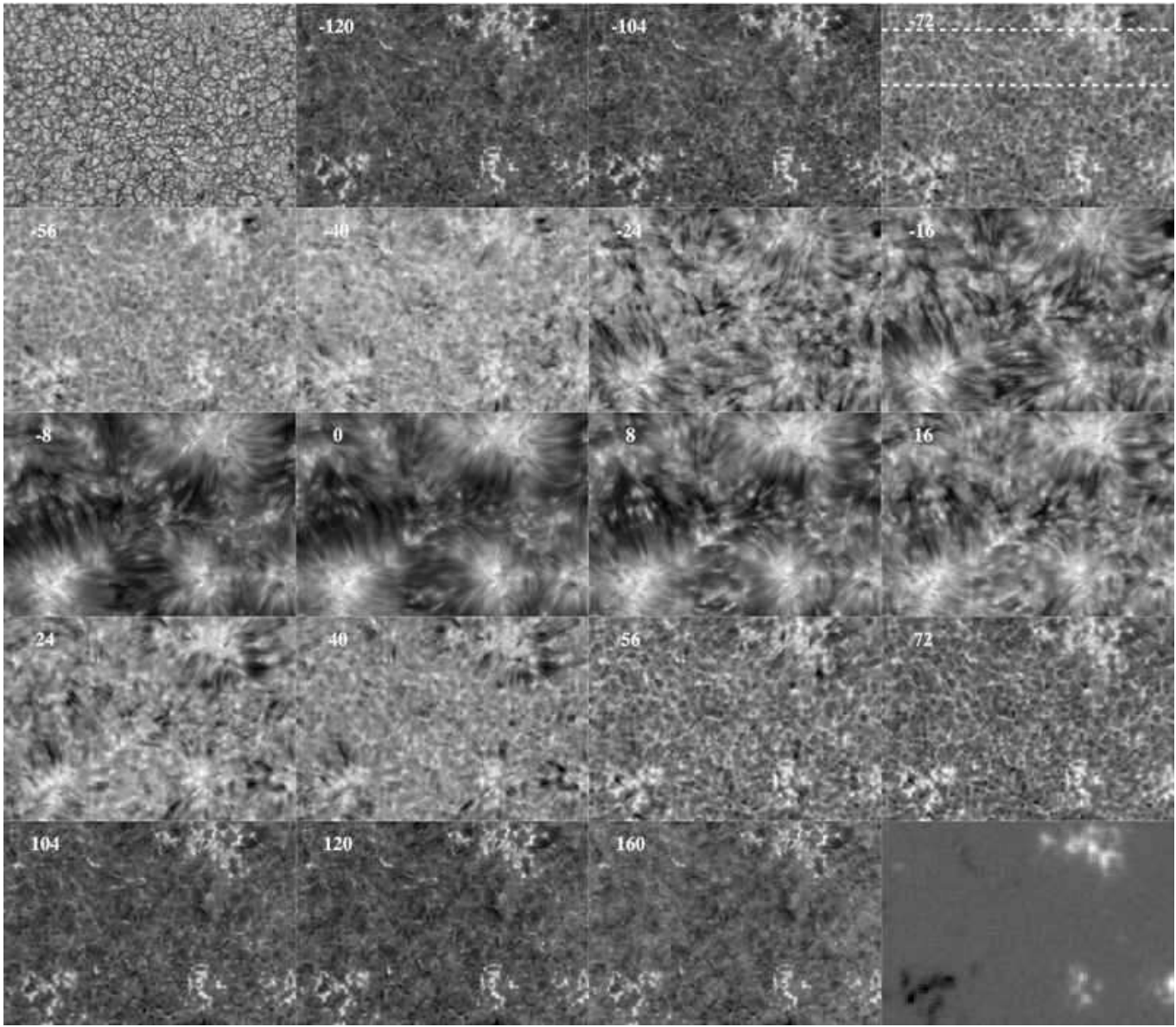}%{20040602.eps}
\caption{Same as Fig. \ref{f_20040531}, for June 2, 2004 data. 
In the speckle-reconstructed broadband image many bright points can be appreciated in the
network areas. The MDI map shows a strong, bipolar network (image scaled between
$\pm$500 G). Note how the fibrillar structures originating from the network elements, and
well visible in the inner wings and core images, extend over a substantial
fraction of the FOV, differently to what occurs in the area depicted in Fig. 
\ref{f_20040531}. The two dashed lines in the upper-right panel indicate the positions where the slit spectra of Fig. \ref{f_spectra} have been obtained.}
\label{f_20040602}
\end{figure*}

IBIS is a tunable narrowband filter, whose
main components are two air-spaced, 50 mm diameter Fabry-Perot interferometers \citep{2006SoPh..236..415C,ibis2}. 
%Cavallini 2006; Reardon & Cavallini submitted
IBIS operates in the spectral range
580--860 nm, and can provide quasi-monochromatic images at
any wavelength in that range by suitable tuning of the interferometers. However, 
due to the periodic nature of the spectral transmission profile, the analysis of
any given spectral line  requires the use of a 
prefilter, of 0.3--0.5 nm FWHM, that isolates the central transmission order. 
Currently a prefilter for the \CaII 854.2 nm line of the IRT is available for IBIS. The filter
was provided by Barr Associates, with a FWHM 
of 0.46 nm, 35\% transmission,
centered at 854.25 nm.  The central wavelength of the filter is slightly offset to the red from
the nominal central wavelength of the \CaII 854.2 nm line to allow the filter to be tilted 
in the beam to avoid reflections. This also permits the passband to be 
minimally tuned in a range of approximately $\pm\,$0.1nm from the central wavelength.
We remark also that in the case of broad lines such as the \CaII IRT the limited prefilter passband 
effectively
prevents the sampling of the line all the way to the continuum, as shown in the upper panel of Fig. 
\ref{f_inst_profile}. 

The instrumental spectral transmission of IBIS has been accurately calculated and the FWHM at 854 nm was determined to be 4.4 pm   \citep{ibis2}. 
%Reardon & Cavallini  submitted 
The spectral purity of the instrumental profile is quite high, with over 95\% of the total
transmission coming from within a range of $\pm\,8$ pm around the peak of the profile
and spectral parasitic light, arising from the repeating secondary 
orders of the transmission profile, of only 1.5\%.
The lower panel of Fig. 
\ref{f_inst_profile} shows how this is accomplished with  the combination of two interferometers and the prefilter, which strongly suppresses the secondary maxima in the Fabry-Perots transmission.

Piezoelectric tuning allows the instrumental profile to be positioned at multiple
wavelengths within a given line, making it possible to acquire full spectral
information over the 80$''$ diameter circular field of view (FOV) of the instrument.  
The \CaII 854.2 nm line is typically sampled at 15--30 wavelength positions, 
depending on the scientific requirements (see following
sections). Given the current rate of acquisition (2.5 -- 4 frames s$^{-1}$), this
translates in an interval of 5 -- 12 s to perform a full sampling of the 
line. A planned upgrade for the camera system in the near future may improve this rate by a 
factor of 2 -- 4. 

In order to obtain meaningful spectral information from sequential images at high spatial resolution,
it is necessary that the location of the structures remain stable during the spectral 
scanning of the line.
The high-order adaptive optics system of the DST \citep{2004SPIE.5490...34R}
%Rimmele SPIE 2004. 
is routinely used with IBIS and greatly aids in stabilizing the image. To remove 
residual image motion, especially at increasing distances from the AO lock point,
we obtain a broadband (``white light'') reference image strictly simultaneously with each narrowband
image. These reference images are obtained through 10 nm wide filters, and during the relatively short time needed for a spectral scan show essentially  the same scene (as compared 
to the spectral images where the structures change dramatically during the scan). They are then
normally used to destretch the images in both channels, or 
for more advanced image reconstruction techniques, such as deconvolution of the narrowband channel image supported by
 MOMFBD \citep{2005SoPh..228..191V}
%van Noort et al., 2005 Solar Physics paper 
or speckle imaging \citep{Woeger_thesis}.

Finally, the spatial scale of the IBIS images is set at $0".082$/pixel.
 With the use of the procedures described
above, it is thus possible to exploit the periods of good to excellent seeing 
occurring at the DST up to the nominal diffraction limit of the telescope
of $\lambda/D = 0".23$ at 854.2 nm.

\section{Observations}\label{s_obs}

\begin{table*}
\caption{Summary of the \CaII 854.2 nm observations described in the paper. Date is
expressed in {\em yyyymmdd}; $\Delta$x is the pixel spatial scale; t$_{exp}$ is the
exposure time; $\delta\lambda$
is the spectral sampling step, while $\Delta\lambda$ is the total sampled range; N
is the number of wavelength points; $\delta$t is the scan time, while $\Delta$t 
is the cadence of the observations. For the case of the \CaII 854.2 nm line, often the sampling is coarser in the extended wings with respect to the core.}
\label{t_data}
\begin{center}
\begin{tabular}{lllllllllll}
Date & Target & Location & $\Delta$x (") & t$_{exp}$ (ms) & $\delta\lambda$ (pm) &$\Delta\lambda$ 
(nm) & N  & $\delta$t (s)& $\Delta$t  (s) & Duration (min)\\
\\ 
\hline
\\
20040531 & quiet Sun & disk center & 0.16 & 25 & 8--16 & [$-0.12$, $0.16$] & 27 & 7
& 19 &  55\\
20040602 & quiet Sun & disk center & 0.16 & 25 & 8--16 & [$-0.12$, $0.16$] & 27 & 7
& 19 & 55 \\
20051001 & small pore & disk center & 0.082 & 50 & 4 &  [$-0.07$, $0.13$] & 35 & 700 & -- & --\\
\end{tabular}
\end{center}
\end{table*}

We present in this paper some example \CaII 854.2 nm data acquired mostly in quiet target areas,
as listed in Table \ref{t_data}.   We remark however that IBIS observations in this spectral line are well 
suited also for studies of solar activity. On the one hand, the 80'' diameter FOV is able to accommodate
large portions of active regions (or even full small ones), thus overcoming a typical problem of spectrographic observations. On the other, the 
ability to obtain spectral profiles in just a few seconds (ten or less) 
allows a detailed study of the dynamics of rapid impulsive events, such as reconnection driven 
explosive events or small flares, occurring in the solar chromosphere.

The sequence of monochromatic\footnote{Due to the classical mounting of the 
interferometers, the spectral passband of IBIS experiences a radial,
wavelength-dependent, blueshift within the FOV. The effect has been removed from all the images displayed in
this paper.}
images of Figure \ref{f_20040531} shows a scan acquired on May 31, 2004. The
actual wavelengths of observation are indicated in the panels as offsets from nominal line core. The target was a
quiet area at disk center, with mostly unipolar, weak network elements within the
FOV. The whole region was positioned at the edge of an equatorial coronal hole, as
seen from  EIT 171 \AA~full disk images.  The first
panel shows the corresponding white light image, while the last one displays a
cotemporal, high resolution MDI map of the longitudinal magnetic flux.

Fig. \ref{f_20040602} shows  the data
acquired on June 02, 2004, with the same setup as May 31, 2004. The target was again a quiet 
area at  disk center, this time roughly encompassing a full
supergranule surrounded by some enhanced, bipolar network. The whole region
appeared as a fading coronal bright point in the EIT 171 \AA~full disk images, where occasional bouts of small scale activity were observed.  Numerous bright points as well as several rapidly evolving
small pores can be
distinguished in the broadband image
in the regions of the network, testifying to a stronger magnetic
flux as seen in the MDI panel. This dataset has been analyzed in
\citet{ 2007A&A...461L...1V} 
%Vecchio et al 2007, A&A letter on shadows;
and in part in \citet{2006A&A...450..365J}. 
%Janssen & Cauzzi 2006

\begin{figure}
\includegraphics[width=5.6cm,height=8.4cm,angle=-90]{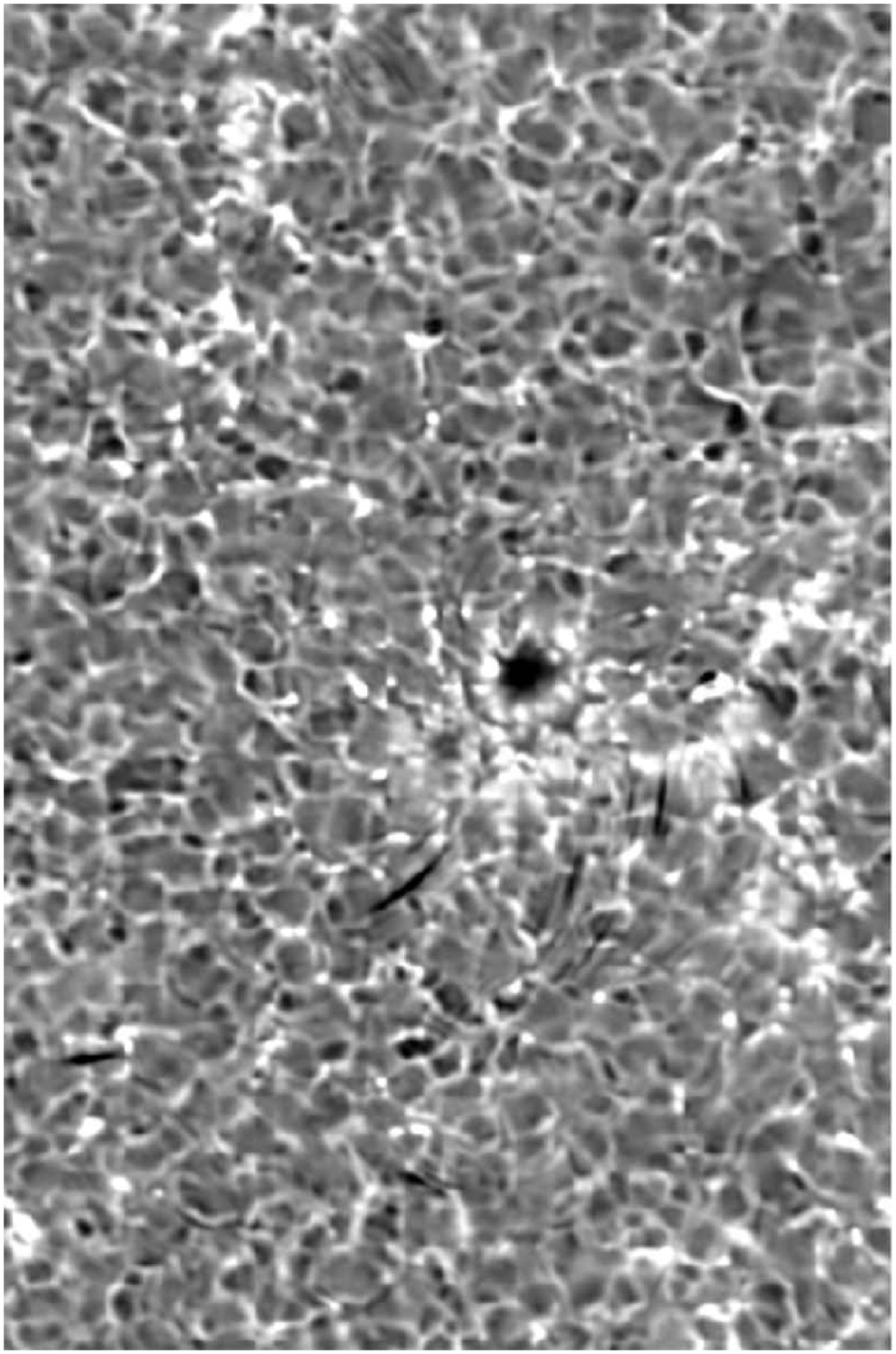}%{nb.recon.00.eps}

\includegraphics[width=5.6cm,height=8.4cm,angle=-90]{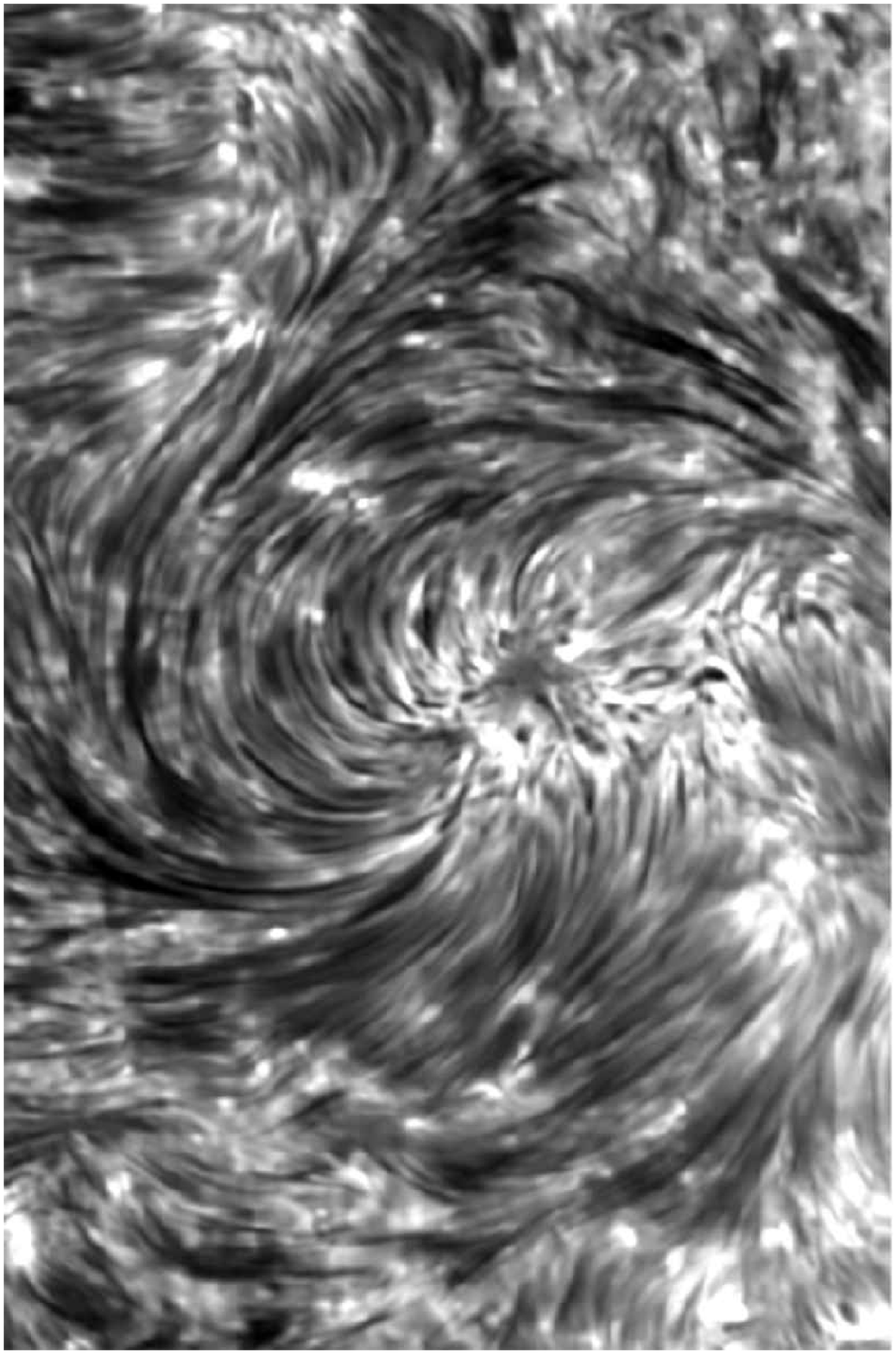}%{nb.recon.11.eps}
\caption{Speckle-reconstructed sample images from the weak plage region of Oct. 1, 2005. Upper panel: image obtained at 65 pm in the blue wing of the line. FOV is about 60''$\times$40''.  Bottom panel: corresponding image at 20 pm in the blue wing, obtained 200 s later. Images have been reconstructed applying the speckle code of  \citet{Woeger_thesis}. See the online version for a movie of these data.}
\label{f_20051001}
\end{figure}

Fig. \ref{f_20051001} shows the target region of Oct. 01, 2005, encompassing a
small pore (3''--4'' diameter) within a region of weak plage near disk center. This dataset was acquired to test the procedure for performing post-facto 
image reconstruction using the speckle code of \citet{Woeger_thesis}, hence   
50 repeated exposures were obtained at each wavelength point, resulting in a long total acquisition time. 
The top panel displays the reconstructed image at about 60 pm in the blue wing, while the bottom one gives the image at 20 pm, again in the blue wing. A movie displaying a cotemporal G-band frame, together with the tuning of
the IBIS passband through the blue wing of the line is available in the online
version of this paper\footnote{Movie 1,  {\rm http://www.arcetri.astro.it/science/solar/IBIS/movies/
CaII.8542.scan.01Oct2005.movie.gif}}. This data clearly illustrates the
wealth of fine structure information the \CaII IRT can provide when observed
at high resolution \citep[see also Fig. 11 in][]{2007ASPC..368...27R}.
%Rutten Coimbra

\section{Line formation} \label{s_line_formation}

The most striking characteristics of the narrowband  sequences of Figs. \ref{f_20040531} -- \ref{f_20051001} is the dramatic change in scenery when the instrumental passband moves from the wings towards the core of the line - the latter displaying swarms of fibrillar structures usually associated only with H$\alpha$ core diagnostics.  The  ``critical wavelength'', at about 30--40 pm from the core, corresponds to the inflection point in the line where the steep inner wings leave place to the much shallower outer wings (see the profiles in Fig. \ref{f_inst_profile}).  

The sharp knees in the line profile mark the transition from
LTE absorption line in the photospheric outer wings to the Non-LTE
chromospheric absorption line core.
This transition is a reflection of two effects, in the line source
function and in the line opacity, respectively.
In the photosphere, where LTE applies, the line source function
decreases with height with the electron temperature, giving rise to
a slow decrease in line wing intensity towards line center. Indeed, because of their LTE formation, the wings of the 854.2 nm line are
used as accurate temperature diagnostics in solar type stars
\citep[e.g.,][]{1979ApJS...41..481L,1987A&A...181..103S}.
%% Linsky et al 1979, Smith & Drake 1987
%
In the chromospheric layers collisional excitation and de-excitation
loses dominance to radiative transitions as the main population
mechanisms in the line.
Photon losses to outer space propagate inward because of the scattering
in the line that results.
With these losses the radiation field in the line can no longer
maintain the upper level population of the line to LTE levels,
and the line source function drops steeply with height.

\begin{figure}[ht]
\includegraphics[width=8.2cm,height=7cm]{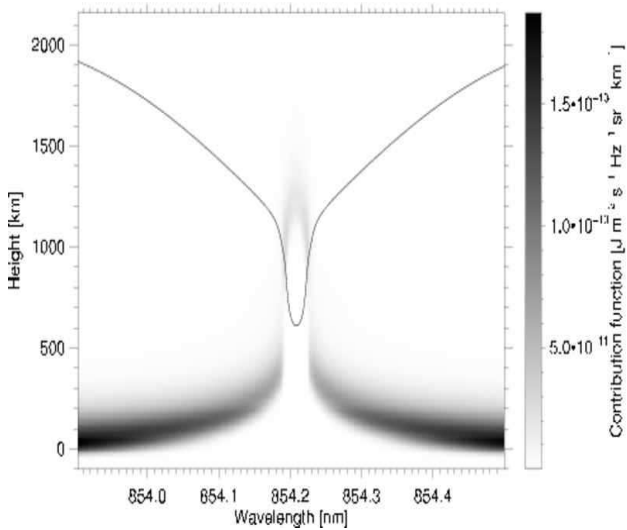}%{con.eps}
\caption{Contribution function for \CaII 8542 computed in plane-parallel, 
hydro-static average quiet-Sun atmosphere
\citep{Fontenla+Avrett+Loeser1993}.
Height zero refers to level at which optical depth in the continuum at 500 nm
is unity.}

\label{f_cf}
\end{figure} 

The sharp transition between these two regimes results 
from a gap in the line opacity around the temperature minimum.
Since the lower levels of the IRT are the metastable
3d $^2$D$_{3/2,5/2}$ levels, their population in low
temperature regions is reduced compared to the ground level
population.
This is a small effect because the excitation energy of the D
levels is only 1.5 eV (e.g. much smaller than the 10 eV of the hydrogen
H$\alpha$ line for which the gap in line opacity is therefore
much more pronounced).
Nevertheless, because of this gap in line opacity, the formation height quickly shifts from photospheric
to chromospheric for small wavelength differences around the knees
of the line.
This quick transition in formation height is obvious
in the line intensity contribution function, displayed
in Figure \ref{f_cf} as a function of wavelength (on the horizontal
axis) and height in the atmosphere (on the vertical axis).
At wavelengths for which the height 
of $\tau_\lambda \sim 1$ corresponds to the height of the line opacity gap,
the contribution function rises steeply with height,
indicating that the line intensity sensitivity quickly changes
from photospheric to chromospheric 
\citep[see
also][]{1989A&A...213..360U,2002ChJAA...2...71Q}.
%Uitenbroek 1989; Qu 2002.

The very narrow spectral passband of IBIS makes it possible to  clearly distinguish these two regimes directly in the imaging, with a rapid change with wavelength in observed morphology.  The outer wing images thus mostly display convective structures, only occasionally ``contaminated'' by  fibrils that show up as  thin dark strikes.  
The core images instead display mostly fibrillar, chromospheric structures,  that appear most evident in the blue or red inner wings depending on their line width and Dopplershift. Only in very quiet portions of the FOV  the fibrils are not the predominant feature, and sporadic bright points with size of the order of 1 arcsecond become visible in the core (see Sect. \ref{s_core}).

\section{Morphology} \label{s_morphology}

\subsection{Outer wings -- quiet photosphere} \label{s_wings_quiet}

As just discussed, the high spatial resolution images obtained with IBIS in the outer wings of the \CaII 854.2 nm line are dominated by photospheric structures. In particular, the pattern of reverse granulation, i.e. a partial reversal of the contrast between granules and intergranules with respect to the continuum \citep[see, e.g.][and references therein]{2004A&A...416..333R},
%Rutten et al 2004, on reverse granulation
stands out very clearly within the quiet portions of the FOV. This is visible, for example, by comparing the white light images of Figs. \ref{f_20040531} and \ref{f_20040602} vs. the panels at $-100$ pm \citep[see also][]{2006A&A...450..365J},
% Janssen & Cauzzi 2006
or, even more clearly,  in the first images of Movie 1. Looking carefully at the wing images of Figs. \ref{f_20040531} and  \ref{f_20040602}, one can also notice a slight asymmetry between the blue and the red wings, with reversed granulation being more prominent in the blue, at equal distance from line core. This is due to the effect of cross-talk between intensity and velocities in the convective structures, much as the case for monochromatic images acquired in the wings of typical photospheric lines \citep{2006A&A...450..365J}.
%Janssen & Cauzzi 2006

As one moves further out in the wings, reverse granulation gradually disappears, and the quiet areas show just a diffuse intensity without any obvious structure  (compare e.g. the last panel of Fig. \ref{f_20040602}, at 160 pm from line core). Given the broader sampling of the line in the red (cf. Table \ref{t_data}), this effect is mostly visible in the red-most wavelengths displayed in the Figures. As mentioned in Sect.  \ref{s_inst}, normal IBIS sampling does not reach continuum wavelengths 
(although technically feasible, the measures would be much noisier and affected by a high level of spectral parasitic light). Hence, our observations do not go far enough from the line core to see the normal granular 
structures of the low photosphere.

Using 3-dimensional radiative hydrodynamics simulations, \citet{2007A&A...461.1163C}
% Cheung et al 2007 Reverse granulation
recently offered the most comprehensive
explanation of the horizontal structure of the photospheric temperature, and of reverse granulation in particular. The latter results from the interplay between cooling of granules rising and expanding within the optically thin layers of the photosphere, and radiative heating acting against it. The (horizontally averaged) layer at which the temperature contrast between granules and intergranules vanishes is positioned at about 150 km above the classical photospheric surface ($\tau_{500}=1$);  photospheric diagnostics originating above this height are bound to show evidence of the reversed temperature contrast with respect to the continuum granulation. 
This is indeed consistent with the wing observations presented above, when interpreted through the contribution function of Fig. \ref{f_cf}: at about 0.1 nm from line core the layers contributing most to the intensity are above $\sim$150 km, i.e. a level where reversed granulation is already present. Instead, at about 0.15 nm the intensity forms slightly lower, in intermediate layers where the granular contrast is much reduced.

\subsection{Outer wings -- the magnetic photosphere} \label{s_wings_mag}

Besides reverse granulation, the outer wing images very clearly display the location of magnetic structures, such as the plage or quiet network visible in Figs. \ref{f_20040531}, \ref{f_20040602} and 
\ref{f_20051001}. Given their large temperature sensitivity, the intensity in the outer wings basically maps the lower opacity, higher temperature small scale magnetic elements.  Their visibility is further enhanced at some wavelengths by the reduced contrast of the solar granulation.
The co-temporal MDI high resolution magnetic maps for the case of May 31, 2004 and June 2, 2004  demonstrate that there is a one-to-one correspondence between bright wing structures and magnetic patches.  Actually, the higher resolution IBIS images clearly
reveal how single magnetic features visible in MDI are composed
of several distinct structures of sub-arcsec size, both in the network and possibly, the internetwork \citep{2003A&A...409.1127J}.

The use of intensity proxies as a diagnostics of small-scale magnetic elements in the photosphere has been recently investigated by \citet{2006A&A...452L..15L}, 
%Leenaarts et al 2006
by numerically synthesizing several spectral diagnostics in a snapshot of a 3D solar magnetoconvection simulation.
They identify the outer wings of \CaII 854.2 nm as one of the best proxies (albeit at the price of a reduced spatial resolution with respect to the wings of H$\alpha$ or H$\beta$), consistently with the results shown here. Somewhat at odds with their results, however, we find that the intensity images at $-$0.09 nm from line core already show the reversed granulation pattern (as discussed in the previous Section), rather than normal granulation.  In our data magnetic structures are hence better identified in intensity images obtained further out in the wings (up to the limit of our sampling, i.e. 0.16 nm from core), where the granulation contrast almost vanishes. This discrepancy should be further investigated by a more
detailed comparison between simulations and high quality observations.

\subsection{Core structures} \label{s_core}

Narrow band \CaII 854.2 snapshots, acquired around the line core wavelength in the quiet regions, display a clear segregation of the FOV in various components (see Figs. \ref{f_20040531} and \ref{f_20040602}).  The distinction is even clearer if one has access to the temporal dimension: a movie in the online version\footnote{Movie 2: {\rm http://www.arcetri.astro.it/science/solar/IBIS/movies/
31May2004.core.gif}} shows the evolution of the nominal \CaII 854.2 nm line core as observed over the full FOV and for  the full duration of the observations of May 31, 2004 (55 minutes with cadence of 19 s). Although the seeing conditions were at times variable, and certainly worse than for the data of June 02, 2004, we chose this dataset as it provides a clearer picture of the quietest internetwork regions. In particular, an almost complete supergranular cell ($\sim$28 Mm in diameter) is visible in the central-right portion of the FOV.  

The first feature that one discerns is the magnetic network,  again outlined by clusters of small, bright elements that faithfully map the presence of photospheric magnetic elements.
The appearance of the network elements in the core images is more diffuse with respect to the wings of the line (Sect. \ref{s_wings_mag}),  owing to the lateral spread of the magnetic field with height.  However, it must be remarked  that their spatial extension seems to be very dependent on the instantaneous seeing conditions, as well visible from the movie. As long known, the chromospheric network elements evolve relatively slowly, with single features persisting for several minutes, and they present very different spectral properties and dynamics than the rest of the FOV (Sect. \ref{s_spectral}).

In contrast, the internetwork portions of the FOV are seething with smaller bright features, rapidly evolving, immersed in an otherwise very dark environment.
These bright points reach the same intensity as the network points (about 1.5--2 times the average core intensity), but are more sharply defined as roughly circular areas with 1''--3'' diameter, and evolve with timescales of 1--2 minutes. At their darkest, they reach about 0.75 times the average core intensity. 
Many of the points seem to reappear throughout the course of the observations, 
with a cadence of a few minutes, and about 10-20 of them are present at any given time within the cell interior. These properties are in complete analogy to the properties of the H$_{2V}$ and K$_{2V}$ ``grains'' \citep[see the review of][]{1991SoPh..134...15R}. The latter have been 
explained as due to shocks generated by acoustic waves, with frequencies  slightly above the acoustic cutoff, originating in  the photospheric layers and propagating upwards in the absence of magnetic fields   \citep{1995ApJ...440L..29C,1997ApJ...481..500C}. 
%Carlsson & Stein 1995, 1997
The radiative-hydrodynamical  model of Carlsson \& Stein indicates that indeed the shocks should be well visible  in quiet areas also in the \CaII 854.2 nm line \citep{2006ApJ...640.1142P} but, to our knowledge, they had never been convincingly observed in this spectral signature.  IBIS observations thus open the way to a more complete study of the phenomenon, as both the spectral and spatial domains are accessible at once, with much statistics provided by large FOVs and  the opportunity to analyze different magnetic topologies. Results of such an analysis will be presented in a forthcoming paper \citep{vecchio_shocks}.

The third, prominent feature in the \CaII 854.2 nm core images is due to fibrils,  that seem to originate from  even the smallest magnetic elements (compare Movie 1). Their appearance is strongly reminiscent of the structures  observed in H$\alpha$ \citep[see the recent high resolution observations of][]{2007ApJ...660L.169R}, although we don't have here any direct comparison between the two spectral signatures. Much as for the H$\alpha$ fibrils, the structures observed in \CaII 854.2 differ sensibly between active regions and quiet Sun. More active regions harbor long, and relatively stable fibrils together with shorter dynamic fibrils ({Fig. \ref{f_20051001}), while the 
quiet Sun displays both long mottles, connecting the stronger magnetic concentrations in the network (a clear example is in the lower half of the FOV of Fig. \ref{f_20040602}) and shorter ones, presumably  outlining field lines that reach outward to  the corona and interplanetary space. The structures visible in Movie 2 most probably belong to the latter category, as the FOV lies at the edge of a coronal hole.

Several of the fibrils characteristics are worth of notice: first of all, they occupy a large fraction of the FOV,  even in the quietest of instances. For example, in the region observed on May 31, 2004, they  occupy between 30 and 50\% of the ``internetwork''  area, while they become the dominant component in the weak plage region of Fig. \ref{f_20051001}. Second, their dynamic is completely different from the internetwork areas: looking at the movie, one is given the clear impression of material flowing from the network elements outwards, i.e. of  predominant {\em transversal} motions with respect to the solar surface, as opposed to the mostly vertical dynamics of the bright points. 
We do not observe significant lateral swaying of the fibrils, although this may be due to the lower 
spatial and temporal resolution compared to \citet{2007ApJ...660L.169R}, who had a 1 s cadence (compared 
to the 19 s in our data)  and a spatial resolution nearing the diffraction limit of the Swedish Solar Telescope (SST, 0.17" at the H$\alpha$ wavelength).
Finally, in some instances, small bright points as those described above are visible within the fibrils; whether because of physical lateral motions, or because of changes in the opacity of the structure, at times the fibrils let us glimpse the quiet atmosphere underneath.

\subsection{Fibrils in \CaII ?} \label{s_fibrils}

The overwhelming presence of fibrils in the \CaII 854.2 nm core images  was an initial surprise of the IBIS observations, as  they had never been observed at this level of detail other than in the H$\alpha$ core and inner wings. Earlier \CaII 854.2 nm spectroheliograms by \citet[][Plates 16--18]{title_66} as well as MSDP observations
%in this line they only have sunspots, and not such a large FOV either. Mottles are studied with Halpha
probably did not reach the necessary spectral and spatial resolution in order to well appreciate these characteristics, although MSDP  data have shown the presence of arch filament systems in emerging flux regions \citep{2000A&A...355.1146M}.
%Mein, Briand et al 2000 

A striking point is that fibrils are normally absent from  \CaII H and K  imaging, unless observed at very high spatial resolution in the core of the lines, when they might appear as bright, upright stalks in network areas \citep[e.g.][]{2006ASPC..354..276R,2007ASPC..368...27R}.
%Rutten, both Sac peak and Coimbra meeting
This has led to the (often unspoken) assumption that the latter lines sample a `different', possibly lower  chromosphere than H$\alpha$,  and that the enhanced visibility of fibrils in H$\alpha$ imaging is due to some peculiarities of the line formation, in particular its dependence on photoionization-recombination mechanisms.  

Given the strong interconnection between \CaII H and K  and the IRT formation discussed in the previous Sections, why then does the \CaII 854.2 line display such a different ``chromospheric'' scenario? We believe that a large part of the problem lies not in the infrared line formation, but in the strong observational limitations that still affect  \CaII H and K  data.  In other words, the issue is not why the \CaII 854.2 nm line displays the chromospheric fibrils, but why the H and K  do  not show them!  Essentially, the chromospheric signal of the \CaII H and K  lines is confined to the rather narrow core  (with Doppler width of 10--15 km s$^{-1}$, i.e. $\le$20 pm at 400 nm),
while most of the imaging is performed with broad filters with passbands in the range 30--100 pm. These however strongly dilute the core signal,
resulting in images that are heavily biased toward the wing, upper-photospheric, signal. Scattered light, and general photometric noise further degrade the signal of interest. Finally, even with these relatively broad filters, the low photon flux near the core of these very deep lines (due also to the low transmission of the filters)
requires long exposures, which compounds the increased seeing-induced distortions at the shorter
wavelengths.
For these reasons only very bright chromospheric structures (for example those related to network points) can survive the smearing and still be distinguishable above the photospheric background in the ``core''  images \citep{2006ASPC..354..276R}.
To  better prove our point, a direct comparison of simultaneous \CaII K filtergrams and \CaII 854.2 IBIS data will be presented in a forthcoming work. Preliminary results are reported in \citet{2007ASPC..368..151R}. 

\begin{figure}[ht]
\includegraphics[width=8.4cm,height=9cm]{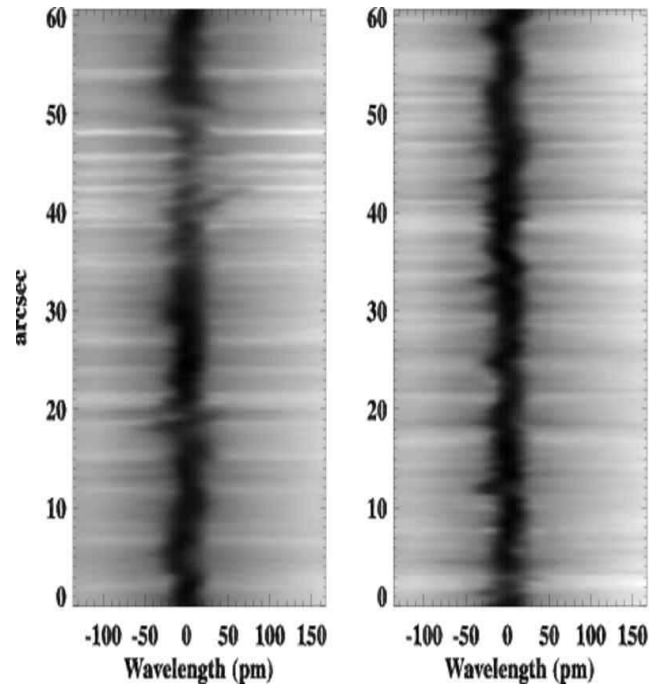}%{slit.eps}
\caption{``Slit'' spectra obtained around the middle of the sequence of June 02, 2004, from two horizontal cuts of the data displayed in Fig. \ref{f_20040602}. The x-axis is the wavelength distance from the average line core position. The left spectrum corresponds to the upper slice, across network  areas,
 while the right one corresponds to the lower slice, and represents quiet internetwork areas.}
\label{f_spectra}
\end{figure}

\section{Spectral structure}\label{s_spectral}

\begin{figure}[ht]
\includegraphics[width=8.4cm,height=14cm]{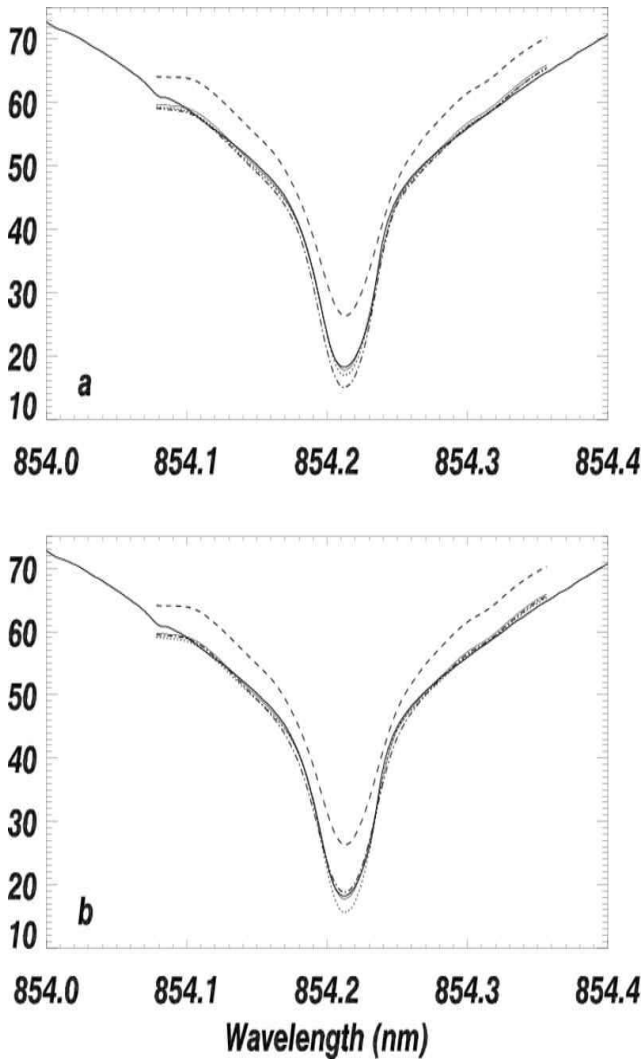}%{average_ref.eps}
\caption{ Average line profiles for the data acquired on June 02, 2004. Panel a: the spectrum averaged over the whole dataset (thin solid line) is compared to the Liege atlas (thick solid). The effect of the prefilter transmission have been removed from the observations. The profiles averaged over different solar features defined by using an intensity threshold are also shown:  network points (dashed); fibrils (dot-dashed) and internetwork (dotted).  Panel b: As panel a, but this time fibrils and internetwork areas were defined using dynamical properties.}
\label{f_avprof}
\end{figure}

The most attractive quality of two-dimensional spectrometers lies in the possibility of obtaining sufficiently detailed spectral information over an extended FOV, in a period short with respect to the evolutionary times of the structures. We describe in this section some characteristics of the \CaII 854.2 nm spectral profiles obtained from the IBIS observations presented above.
We use the quiet Sun dataset obtained on June 2, 2004, that can be easily compared to earlier results. 

\subsection{Spectra and average line profiles}\label{s_lineprofile}

Fig. \ref{f_spectra} 
shows two ``slit spectra'' derived from two cuts along the horizontal direction of the data displayed in Fig. \ref{f_20040602}, as indicated in the Figure. The spectral axis  is obtained interpolating the sequence of  monochromatic values for each pixel along the cut. Given the acquisition sequence, the time also runs with the wavelength, and an interval of about 7 s separates the blue-most from the red-most  wavelengths. The left spectrum 
intersects several network points 
(i.e. between positions 40 and 50 along the slit) 
and fibrils, while the right one 
covers mostly quiet areas. The quality of the spectra is high, and appears comparable to that of data acquired at high spatial resolution with normal spectrographs (see for example the spectra obtained in the \CaII 866.2 nm line at the Swedish Solar Telescope by Langangen et al. 2007). Numerous indications of the dynamics are visible in the two regions, but with very different characteristics. The left panel displays strong redshifts of the line in several pixels (e.g. around positions 40 and 52, that become progressively stronger within a couple of arcsec, and probably involve dynamic fibrils \citep{2007ApJ...655..624D}.
%De Pontieu et al. 2007, ApJ 655
In the right panel, instead, stronger blueshifts are visible  (e.g. position 12, 25, 33 along the slit), that are less extended in the spatial direction and have more constant amplitude. These are related to the development of acoustic shocks in non-magnetic areas  \citep[][]{2006ApJ...640.1142P,vecchio_shocks}.
%Pietarila et al 2006, Vecchio et al 2007 submitted

 In Fig. \ref{f_avprof} we
show a comparison between the Liege atlas (thick solid line) and the average observed line profile, obtained summing over the whole dataset (more than $2\times 10^7$ individual profiles, thin solid line). The effects of the prefilter transmission (cf. Fig. \ref{f_inst_profile}) have been accounted for, and the wavelength offset has been adjusted so that the minimum of the average profile coincides with the atlas line core position. 
The general agreement between the profiles is excellent, showing discrepancies of less than 2\% of the continuum intensity around the inflection point in the wings (the larger difference at the blue limit of the observations is due to a telluric line). Differences of this amplitude are probably unavoidable, due both to slight residuals in the data calibration and to the intrinsic variability of the solar structures in time and space.
Still, all the general characteristics of the atlas profile are well reproduced, including the asymmetry of the line core typical of chromospheric lines \citep[``inverse C-shape'',][]{2006ApJ...639..516U}.
%Uitenbroek 2006

What kind of individual profiles enter the average? To gain insight into this issue, we have attempted a spatial classification for the three features discussed before: network, fibrils, and quiet internetwork. In this kind of analysis, often an intensity threshold is adopted in the definition of separate classes of pixels \citep[e.g.][]{1990A&A...228..506D}. However, as clear from Sect. \ref{s_core}, the chromospheric scene is probably too complex to be properly described by a single, static parameter. In particular,
the fibrils, very prominent at the core wavelengths, appear here both dark or bright with respect to the average contrast, making their recognition quite difficult.  As a first attempt we have identified the three classes of pixels by adopting an intensity threshold applied to the temporal average of the observations, but using, for each feature, the intensity at the wavelength positions where they are most clearly separated from the background. First, the network has been defined as the bright pixels  in the far wing of the line (at $+$155 pm from line core, having intensity above average  plus  one time the r.m.s.). This threshold clearly identifies all of the photospheric magnetic features, as seen e.g. in the cotemporal MDI magnetograms, that also have an obvious correspondence in the chromospheric line core intensity.  Then, the fibrils and the internetwork areas were respectively defined as the dark and intermediate pixels in the inner wing intensity images ($\pm$ 20 pm from line core), using the intensity values of  the average plus or minus 1/3 of the r.m.s. as our limits. These values separated clearly  the two classes of pixels; in particular, they assured that there were no quiet pixels in the vicinity of network points, where fibrils supposedly originate. From our definition, the fractional area occupied was 9\% (network points), 39\% (fibrils) and 35\% (internetwork), with the remaining pixels not assigned to any category. As expected, these latter unclassified pixels lie mostly around the magnetic elements. It is interesting to note that in this dataset the fractional area covered by fibrils is actually larger than the quiet Sun's, reflecting the closed magnetic topology of the region.

The averaged line profiles for the three classes are displayed in Fig. \ref{f_avprof} as well. The network points average profile (dashed line) is much brighter than the reference one at all wavelengths (however note that we don't have any information on the continuum values), is slightly redshifted (about 150 m s$^{-1}$),  and has a core asymmetry less pronounced than the average profile. The fibrils' profile (dash-dotted) is almost complementary to the network one: it is slightly blueshifted with respect to the average over the whole FOV (about 100 m s$^{-1}$); it is deeper, with a core intensity about 3\% lower (in units of the continuum intensity), while the overall wings remain close to the  average. These same properties have been reported by  \citet{2005AGUSMSP41B..05H} in a study of ``calcium circumfaculae'' using \CaII 854.2 SOLIS data around active regions. He demonstrated how the areas of deeper, blue-shifted \CaII 854.2 profiles coincided with the extended  fibrilar regions observed in H$\alpha$ core. Finally, the internetwork profile (dotted line) is only slightly deeper than the average, while most of the other profile characteristics remain very similar to the average over the whole FOV.

\subsection{Velocity distribution}\label{s_velocity}

From the spectral profiles described above, we have calculated line-of-sight velocities for each pixel in
the FOV and each time step. To this end we used the Doppler shifts of the intensity minima of the spectral profiles, derived from second degree polynomial fits of the line core. The zero has been defined as the spatio-temporal average position over  the whole dataset.
This approach has often been used for the case of quiet chromospheric dynamics  \citep[e.g.][]{1990A&A...228..506D},
%Deubner & Fleck 1990
but  one must be aware that the derived values  might not be fully representative of the actual plasma velocities, especially in the presence of non linear phenomena such as shocks. In particular, we expect that our velocity determination might fail for a certain fraction of the pixels in the magnetic areas  (2--3\% of the total pixels), for which we observe emission in the blue wing, similarly to the case reported in \citet{2007ApJ...663.1386P}.  However, since we are only interested here in giving general properties, and given the overall small number of such pixels, we will keep this definition as a convenient parametric description of the dynamics of the system.

The upper panel of Fig. \ref{f_velocity} displays the distribution of velocity values over the whole dataset (thick solid line). Positive values indicate red-shifts, i.e. downward motions. The distribution is skewed towards positive values,  justifying the peak at the negative (upward) value of $-0.12$ km s$^{-1}$ since we defined the zero as the average value over the whole dataset. About 97\% of the pixels are contained in the $\pm$ 3 km s$^{-1}$ range, but outliers can reach high velocity values, above 10  km s$^{-1}$. At these larger values the distribution becomes markedly asymmetric in the positive quadrant.  

Decomposing the distribution into the three types of solar features described above, i.e. network (dashed line), fibrils (dash-dotted) and internetwork areas (dotted), one sees that the various distributions have different widths, with the fibrils displaying the least number of extreme vertical flows, especially lacking strong downflows. The r.m.s. values of the distributions are, 1.5, 1.4 and 1.25 km s$^{-1}$ for, respectively, network, internetwork areas, and fibrils. Further,
the enhanced positive lobe in the total velocity distribution is largely due to the network points, that account for about 40\% of this signal despite representing less than 10\% of the total pixels. Since neither fibrils nor quiet areas contribute much at the highest redshifts,  the signal missing  with respect to the whole FOV distribution must be due to the pixels left unclassified, that are positioned around the network. This further underlines the highly dynamic nature of the magnetic network elements, that for this enhanced flux region might be related to the dynamic fibrils often observed in plages \citep{2006ApJ...647L..73H,2007ApJ...655..624D}.  

\begin{figure}
\includegraphics[width=8.4cm,height=6cm]{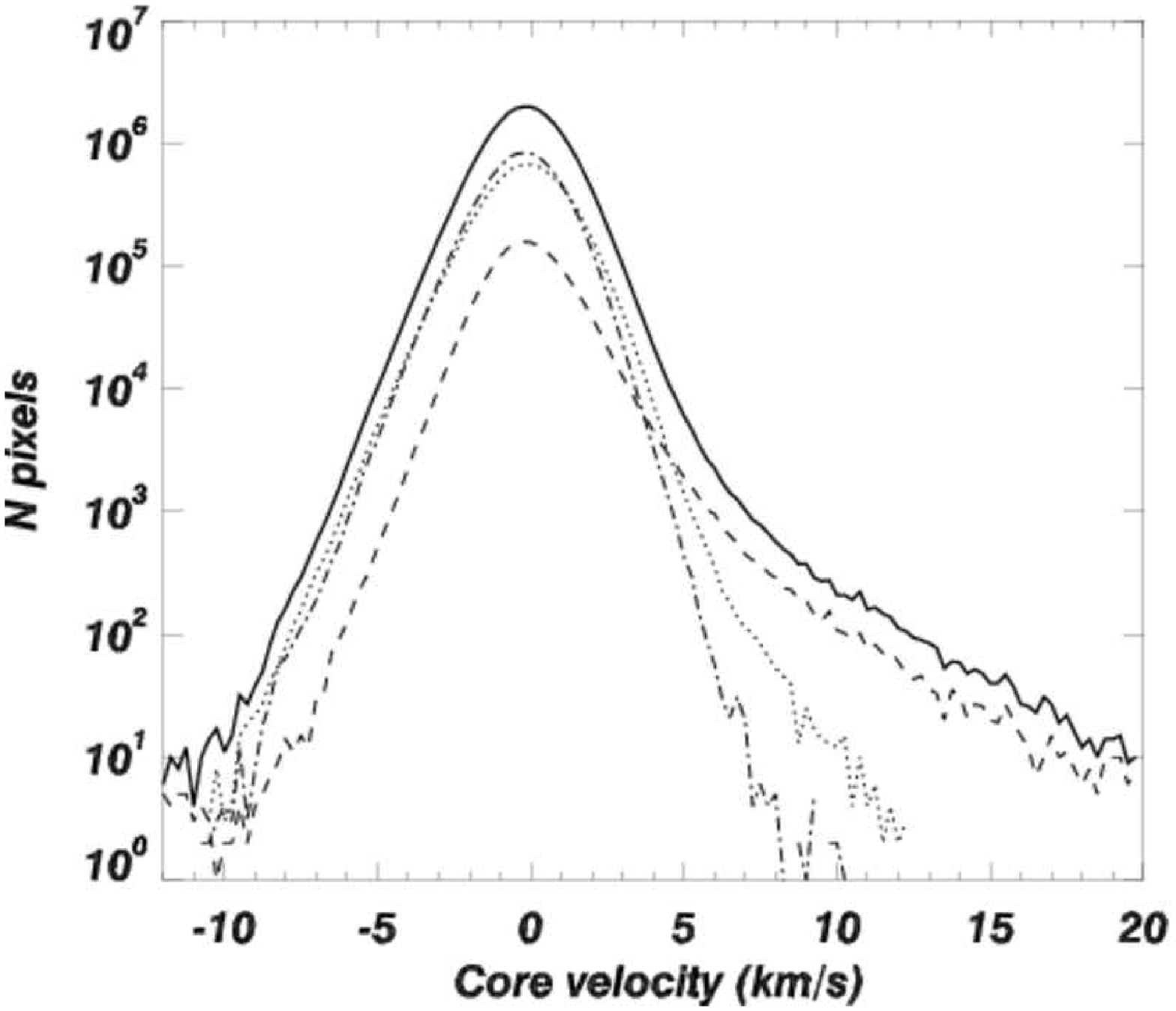}%{hist_vel_int.eps}

\includegraphics[width=8.4cm,height=6cm]{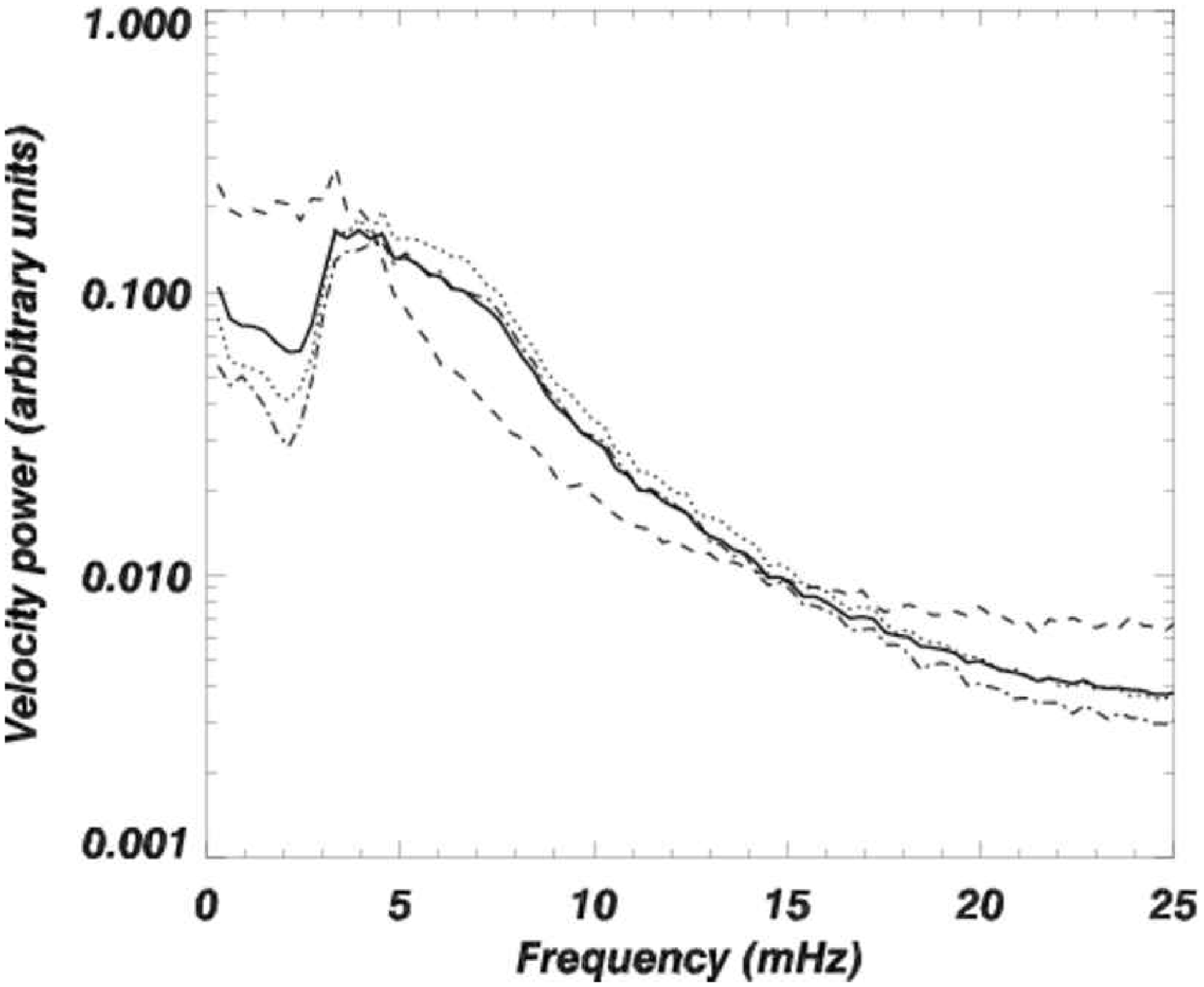}%{pow_vel_int.eps}
\caption{ Upper panel: line-of-sight velocity distribution for the dataset of June 2, 2004. Solid line: overall FOV and time (more than 2$\times 10^7$ values); dashed, dash-dotted and dotted: distributions for, respectively, network points, fibrils, and internetwork. Bottom panel:  velocity power spectra for the same features (symbols as upper panel). The curves represent an average over the corresponding pixels.}
\label{f_velocity}
\end{figure} 

\subsection{Velocity power spectra}\label{s_power}

To address the presence and relevance of periodicities in the \CaII 854.2 nm  dynamics, the temporal evolution of line-of-sight velocities has been investigated via a standard Fourier analysis, performed separately on each spatial pixel. The Nyquist frequency of the observations is 26 mHz, and the frequency step is 0.33 mHz. The thick solid curve of Fig. \ref{f_velocity}  (bottom panel) shows
the resulting spatially averaged velocity power spectrum. It has a broad peak between $\sim$3 and 7 mHz (periodicities between 2.4 and 5 minutes),  that indicates an almost equal contribution of both typical photospheric and chromospheric oscillatory signals to the average. This is fully in agreement with earlier results obtained using single-slit spectrographic observations on quiet solar regions
\citep[e.g.][]{1967IAUS...28..293N,1989A&A...224..245F}. 
%Noyes 1967, Fleck & Deubner 1989
We note that the noise levels, determined by the accuracy of the core fitting procedures, is very comparable between these
IBIS data and previous spectroscopic studies (this accuracy was only achieved through the
use of destretching techniques to correct for differential image motion).

The power spectra relative to the three different atmospheric components defined above are also displayed in Fig. \ref{f_velocity}, using the same symbols as Fig. \ref{f_avprof}. The curves are an average over the corresponding pixels, i.e. to recover  the global power spectrum they must be weighted with their relative occurrence. They confirm several of the properties derived by earlier works, in particular that low frequency oscillations are more prominent in network elements than quiet areas, while  periodicities of 3 minutes or shorter are essentially due to the quiet internetwork \citep[see e.g. the Introduction of][] {1990A&A...228..506D}. 
%Deubner & Fleck 1990 , 
The network points display a clear peak at $\sim$3.5 mHz, but we cannot obviously identify any further maxima at lower frequencies \citep[as claimed e.g. by][]{1997ApJ...486L.145K}. The network power (per pixel element) at the 3.5 mHz frequency is enhanced by about a factor of 2  with respect to the rest of the FOV, as reported in \citet{1993ApJ...414..345L}. 
%Lites Rutten & Kalkofen 1993
Known for some time, these latter properties have recently been related to the leakage of the dominant photospheric  $p$-modes into higher atmospheric layers due to the  lowering of the acoustic cutoff frequency in magnetic elements \citep{2006ApJ...648L.151J,2006ApJ...647L..73H,2007A&A...461L...1V}. 
%Jefferies et al 2006; Hansteen et al 2006, Vecchio et al 2007a

Finally, from the plots of Fig. \ref{f_velocity} we see that fibrils share much of the same properties of the quiet internetwork, with a somewhat reduced power at all frequencies (most evident between 5 and 8 mHz). Although in line with the results of \citet{1990A&A...228..506D}, this is a somewhat surprising result, given the very different properties of fibrils with respect to the ``quiet'' chromosphere, both in terms of morphology and dynamics (Sect. \ref{s_core}). This would also appear to be in contradiction with the results of \citet{2007A&A...461L...1V} which,  taking full advantage of the spatial resolution afforded by these data, showed  how the fibrils are well correlated with the ``magnetic shadows'',  i.e. areas of reduced velocity power at frequencies above the acoustic cut-off, surrounding even the smallest magnetic elements \citep{2001ApJ...554..424J}. Armed with this knowledge, we show in panel b of Fig. \ref{f_avprof} the same kind of average spectral profiles displayed in Fig. \ref{f_avprof} a), but this time the  ``fibrils'' and ``internetwork''  areas were defined by classifying the spatial pixels on the basis of their oscillatory power rather than intensity. Over the frequency range $5.5-8.0$ mHz, all pixels with a velocity power smaller than the mean of the whole FOV were defined as fibrils;  pixels with power greater than the average plus one r.m.s. were defined as quiet. These values clearly distinguish between the two different dynamical regimes. While a slight blueshift (about 70 m s$^{-1}$) is still present in the fibrils' profile, its core intensity now is very close to the general average. On the contrary, the internetwork profile becomes sensibly darker, as could be expected from the visual appearance of the quiet (i.e. non-magnetic) areas in Movie 1. We thus conclude that the adoption of a simple threshold of average intensity in order to discriminate between chromospheric structures can be only of limited value, as it does not effectively separate areas with significantly distinct chromospheric dynamics.

\section{Summary and Conclusions} \label{s_conclusion}

We have presented  an overview of the quiet solar chromosphere as observed in the \CaII 854.2 nm line at high spatial, spectral and temporal resolution. This was made possible through the use of an advanced instrument, the Interferometric BIdimensional Spectrometer \citep[IBIS,][]{2006SoPh..236..415C}, installed at the NSO/Dunn Solar Telescope. To our knowledge, this is the first time that observations of this type and quality are reported for any chromospheric Calcium line.

The stability and large throughput of the instrument, combined with the excellent performance of the high-order adaptive optics system available at the DST, make it possible to investigate the highly dynamic chromospheric environment while maintaining excellent spatial and spectral resolution, despite the obvious drawback (common to all filter-based instruments) of a sequential spectral acquisition. 
A comparison of the average properties of the spectral line profiles and of the global dynamics over a quiet target area with earlier results, obtained with classical fixed-slit spectrographic techniques, demonstrates indeed that the spectral information provided by IBIS is fully reliable and
achieves a quality on par to those of dispersion-based measurements.

The monochromatic images acquired with the very narrow spectral passband and high spatial resolution of IBIS (Figs. \ref{f_20040531}--\ref{f_20051001}) display with amazing clarity the  ``dual nature'' of the \CaII 854.2 nm line. In fact, much as for the \CaII H and K resonance lines, the broad wings of the \CaII 854.2 nm gradually sample the solar photosphere, and provide a clear temperature diagnostics for these layers \citep{1972SoPh...25..357S,1974SoPh...39...49S}. The core is instead a purely chromospheric indicator.
Narrow band imaging highlights these characteristics by showing a dichotomy of small scale features at the different wavelengths. The
outer wings clearly display phenomena of convective origin (such as the onset of  reversed granulation) and the presence of bright magnetic elements. On the contrary, the core images show highly dynamic small scale bright ``grains''  akin to the H$_{2V}$ and K$_{2V}$ grains, and fibrillar structures originating in small scale magnetic structures and outlining the  local magnetic topology. The nearly ubiquitous presence of these structures, particularly around the network magnetic elements, indicates that even at these lower chromosphere heights, the atmosphere is already highly structured by the pervasive magnetic fields, consistent with the structures commonly seen in H$\alpha$. The relative lack of fibrilar structures in observations of the \CaII
H and K line images is probably due to the broader filters typically used (30--300 pm FWHM, the latter being the width of the filter used on {\it Hinode}), which tend to mix the signal from vastly different regions of the atmosphere.

The demonstrated high spectral purity, imaging capabilities, and temporal stability of IBIS open up many possibilities in terms of investigation of the quiet and active solar atmosphere with the \CaII 854.2 nm line. For example, the high temporal cadence at which full spectral profiles can be sampled, and the availability of an extended FOV, make this instrument a very promising tool for the study of small scale activity such as reconnection driven explosive event, or small flares. 

In the case of the quiet Sun,
given the temperature sensitivity of the outer wings of the \CaII 854.2 nm, it would be possible to derive the precise run of temperature with height over an extended FOV, as a check of the realism of 3-D hydrodynamical simulations of surface convection in the upper photosphere. Even more interesting, such data should be well suited to investigate the presence and relevance of waves at different layers in the photosphere \citep{2004A&A...416..333R,2006A&A...450..365J,2006A&A...459L...9W}.
%Rutten et al 2004; Janssen & Cauzzi 2006, Woeger et al 2006 A&A letter
%
Information on the spectral behavior of the line core,  acquired at high spatial resolution over extended periods of time, should be able to shed light on several unresolved issues related to the chromospheric structure, such as the occurrence and relevance of acoustic shocks. While it is commonly accepted that acoustic waves at frequencies above the acoustic cutoff produce shocks in the  quiet chromosphere, 
it is still hotly debated if and how much this kind of waves can contribute to the general chromospheric heating \citep{2005Natur.435..919F,2007ASPC..368...93W}, 
%Fossum & Carlsson Nature 2005, Wedemeyer reply in Coimbra
and what role they play in the overall chromospheric structuring \citep{2007ASPC..368...81A}. 
% Avrett SUMER paper Coimbra 2007
Even more pressing would be an understanding  of the role of acoustic waves and shocks in and around magnetic elements, especially those composing the ``quiet''  network. Several recent papers \citep{2004Natur.430..536D,2006ApJ...647L..73H,2006ApJ...648L.151J} have in fact addressed the possibility that small scale magnetic elements might  channel part of the powerful, low-frequency photospheric $p$-modes into higher atmospheric layers where they can develop into shocks and structure the surrounding environment. As described in the Introduction, the use of a 2-D spectrometer in this case offers a great advantage over fixed slit spectrographs, both because of the possibility to follow {\it a posteriori} the wandering of the magnetic elements during the course of the observations, and of the availability of an extended FOV. The latter supplies a greatly improved statistics as well as important information on the local magnetic topology.
 
In forthcoming papers using IBIS \CaII 854.2 nm observations we will demonstrate that acoustic shocks are a  pervasive characteristics of the quiet chromosphere, even at the relatively low layers sampled by this line  \citep{vecchio_shocks}, and that indeed there is spectral evidence of strong shocks of acoustic origin around quiet magnetic network elements (Cauzzi et al., in preparation). In both cases the presence of the ubiquitous fibrils discussed in Sections \ref{s_core} and \ref{s_fibrils} proves to be a key ingredient of the chromospheric structure, outlining the magnetic canopy and defining atmospheric {\em volumes} with very different dynamical properties \citep{2007A&A...461L...1V}.
%Vecchio et al 2007 A&A Letter
%
This strongly supports the remarks of \citet{2007ASPC..368...27R}, that a full understanding of the chromosphere cannot develop without addressing its  3-dimensional nature. A comprehensive approach is needed, that includes forward 2-D and 3-D numerical simulations coupled with detailed radiative transfer on the one side, and high resolution (temporal, spatial and spectral)  observations over extended fields of view on the other. The latter task is currently best addressed by high performance imaging devices adopting rapidly tunable Fabry-Perot  interferometers such as IBIS,  that eventually will be able also to provide full spectro-polarimetric observations.

Indeed,  although we haven't addressed directly the issue in this paper,  the \CaII 854.21 nm line is likely the most reliable spectral line for
measuring magnetic field strength and orientation in the solar
chromosphere. In fact, its contribution function (Fig. \ref{f_cf}) suffers much less from
the formation gap around the temperature minimum than that of H$\alpha$. As a result, 
response functions
of the 854.2 Stokes V profile to perturbations in the magnetic
field \citep{2006ASPC..354..313U} indicate that Zeeman polarization in
the line would be much more sensitive to the chromospheric field,
and over a more coherent range of heights than  for H$\alpha$ 
\citep{2004ApJ...603L.129S}. Moreover, because hydrogen
is a much lighter element, and has a large Doppler width, the
latter line has intrinsically low linear polarization from the
transverse Zeeman effect, which is proportional to the square
of the ratio of Zeeman splitting over Doppler width \citep[][p. 259]{1994ASSL..189.....S}. 
The CaII H and K lines are less suitable for magnetometry
because they require modeling with partial frequency redistribution
(PRD), and the precise interaction of PRD and Zeeman polarization is
theoretically not well established. The NaI D lines provide magnetic
sensitivity only in the upper photosphere, lower chromosphere
\citep{2006ASPC..354..313U}.

Interpretation of polarized spectra from the the CaII 854.2 nm line, as well
as other chromospheric signatures, will require full inversion techniques, because
of the wide range in formation heights over the width of the lines.
One such a technique has been explored to determine the chromospheric
magnetic field in sunspots  \citep{2000ApJ...530..977S,
2005ApJ...631L.167S} and in the quiet Sun \citep{2007arXiv0707.1310P}, showing a promise that with high resolution spectro-polarimetric
imaging such as those provided by IBIS significant progress can be made in the
determination of chromospheric magnetic fields.}

\acknowledgements{We want to remember here our friend and colleague B. Caccin, 
prematurely deceased in June 2004, with whom we exchanged many discussions about
the adoption of \CaII 854.2 
nm in IBIS. His wisdom and insight are deeply missed.
This paper benefitted from discussion with V. Andretta, B. Fleck, M. Carlsson, R. J. Rutten and A. Tritschler. The continuous help and patience of the observers M. Bradford, J. Elrod and D. Gilliam at the DST are greatly appreciated.
IBIS has been built by INAF/Osservatorio Astrofisico di Arcetri with contributions from the Universit\`{a} di Firenze and the Universit\`{a} di Roma ``Tor Vergata''. IBIS construction and operation has been supported by the Italian Ministero dell'Universit\`{a} e della Ricerca (MUR) as well as the Italian Ministry of Foreign Affairs (MAE). NSO is operated by the
Association of Universities for Research in Astronomy,  Inc. (AURA), under
cooperative agreement with the National Science Foundation. This research
was supported through the European Solar Magnetism Network (ESMN,
contract HPRN-CT- 2002-00313), grant PRIN 2004 MIUR and the Italian Ministry of Foreign Affairs (MAE).}

\begin{appendix}
\section{Captions for the online movies}
{\bf Movie 1} ({\rm http://www.arcetri.astro.it/science/solar/IBIS/movies/
CaII.8542.scan.01Oct2005.movie.gif}): IBIS spectral scan within the blue wing of the \CaII 854.2 nm line.  FOV is about 60''$\times$40'', with spatial scale of 0.082"/pixel. The first image shows the solar scene acquired with a co-temporal G-band channel, and displays normal granulation around two small pores and a weak plage region. The spectral images span from 65 pm in the blue wing of the \CaII 854.2 to line core, in steps of 4 pm. The movies then fades from the core image back to the G-band image. Each monochromatic image has been obtained via speckle-reconstruction \citep{Woeger_thesis} of a burst of 50 images acquired within  20 s. 
The sequence of narrowband images illustrates how sampling the line from wing to core one gradually samples higher layers of the atmosphere, from the upper photosphere (where reversed granulation appears obvious) to the chromosphere dominated by fibrillar structures.
  
\noindent{\bf Movie 2} ({\rm http://www.arcetri.astro.it/science/solar/IBIS/movies/
31May2004.core.gif}):
Temporal evolution of the line core intensity of \CaII 854.2 nm, for the data of May 31, 2004. The spatial scale is in arcsec. The movie spans 55 minutes, with a time step of 19 s (175 images). See text for discussion. 

\end{appendix}


\begin{thebibliography}{73}
\expandafter\ifx\csname natexlab\endcsname\relax\def\natexlab#1{#1}\fi

\bibitem[{{Andretta} {et~al.}(2005){Andretta}, {Bus{\`a}}, {Gomez}, \&
  {Terranegra}}]{2005A&A...430..669A}
{Andretta}, V., {Bus{\`a}}, I., {Gomez}, M.~T., \& {Terranegra}, L. 2005, \aap,
  430, 669

\bibitem[{{Avrett}(2007)}]{2007ASPC..368...81A}
{Avrett}, E.~H. 2007, in Astronomical Society of the Pacific Conference Series,
  Vol. 368, The Physics of Chromospheric Plasmas, ed. P.~{Heinzel},
  I.~{Dorotovi{\v c}}, \& R.~J. {Rutten}, 81--+

\bibitem[{{Beckers} {et~al.}(1972){Beckers}, {Mauter}, {Mann}, \&
  {Brown}}]{1972SoPh...25...81B}
{Beckers}, J.~M., {Mauter}, H.~A., {Mann}, G.~R., \& {Brown}, D.~R. 1972,
  \solphys, 25, 81

\bibitem[{{Carlsson} \& {Stein}(1995)}]{1995ApJ...440L..29C}
{Carlsson}, M. \& {Stein}, R.~F. 1995, \apjl, 440, L29

\bibitem[{{Carlsson} \& {Stein}(1997)}]{1997ApJ...481..500C}
{Carlsson}, M. \& {Stein}, R.~F. 1997, \apj, 481, 500

\bibitem[{{Cauzzi} {et~al.}(2007){Cauzzi}, {Reardon}, {Vecchio}, {Janssen}, \&
  {Rimmele}}]{2007ASPC..368..127C}
{Cauzzi}, G., {Reardon}, K.~P., {Vecchio}, A., {Janssen}, K., \& {Rimmele}, T.
  2007, in Astronomical Society of the Pacific Conference Series, Vol. 368, The
  Physics of Chromospheric Plasmas, ed. P.~{Heinzel}, I.~{Dorotovi{\v c}}, \&
  R.~J. {Rutten}, 127--+

\bibitem[{{Cavallini}(2006)}]{2006SoPh..236..415C}
{Cavallini}, F. 2006, \solphys, 236, 415

\bibitem[{{Cheung} {et~al.}(2007){Cheung}, {Sch{\"u}ssler}, \&
  {Moreno-Insertis}}]{2007A&A...461.1163C}
{Cheung}, M.~C.~M., {Sch{\"u}ssler}, M., \& {Moreno-Insertis}, F. 2007, \aap,
  461, 1163

\bibitem[{{Chmielewski}(2000)}]{2000A&A...353..666C}
{Chmielewski}, Y. 2000, \aap, 353, 666

\bibitem[{{De Pontieu} {et~al.}(2004){De Pontieu}, {Erd{\'e}lyi}, \&
  {James}}]{2004Natur.430..536D}
{De Pontieu}, B., {Erd{\'e}lyi}, R., \& {James}, S.~P. 2004, \nat, 430, 536

\bibitem[{{De Pontieu} {et~al.}(2007){De Pontieu}, {Hansteen}, {Rouppe van der
  Voort}, {van Noort}, \& {Carlsson}}]{2007ApJ...655..624D}
{De Pontieu}, B., {Hansteen}, V.~H., {Rouppe van der Voort}, L., {van Noort},
  M., \& {Carlsson}, M. 2007, \apj, 655, 624

\bibitem[{{Deubner} \& {Fleck}(1990)}]{1990A&A...228..506D}
{Deubner}, F.-L. \& {Fleck}, B. 1990, \aap, 228, 506

\bibitem[{{Eddy}(1973)}]{1973SoPh...29...23E}
{Eddy}, J.~A. 1973, \solphys, 29, 23

\bibitem[{{Fleck} \& {Deubner}(1989)}]{1989A&A...224..245F}
{Fleck}, B. \& {Deubner}, F.-L. 1989, \aap, 224, 245

\bibitem[{{Fleck} {et~al.}(1994){Fleck}, {Deubner}, {Hofmann}, \&
  {Steffens}}]{1994chdy.conf..103F}
{Fleck}, B., {Deubner}, F.-L., {Hofmann}, J., \& {Steffens}, S. 1994, in
  Chromospheric Dynamics, ed. M.~{Carlsson}, 103--+

\bibitem[{Fontenla {et~al.}(1993)Fontenla, Avrett, \&
  Loeser}]{Fontenla+Avrett+Loeser1993}
Fontenla, J.~M., Avrett, E.~H., \& Loeser, R. 1993, \apj, 406, 319

\bibitem[{{Fossum} \& {Carlsson}(2005)}]{2005Natur.435..919F}
{Fossum}, A. \& {Carlsson}, M. 2005, \nat, 435, 919

\bibitem[{{Hansteen} {et~al.}(2006){Hansteen}, {De Pontieu}, {Rouppe van der
  Voort}, {van Noort}, \& {Carlsson}}]{2006ApJ...647L..73H}
{Hansteen}, V.~H., {De Pontieu}, B., {Rouppe van der Voort}, L., {van Noort},
  M., \& {Carlsson}, M. 2006, \apjl, 647, L73

\bibitem[{{Harvey}(2005)}]{2005AGUSMSP41B..05H}
{Harvey}, J.~W. 2005, AGU Spring Meeting Abstracts, B5+

\bibitem[{{Janssen} \& {Cauzzi}(2006)}]{2006A&A...450..365J}
{Janssen}, K. \& {Cauzzi}, G. 2006, \aap, 450, 365

\bibitem[{{Janssen} {et~al.}(2003){Janssen}, {V{\"o}gler}, \&
  {Kneer}}]{2003A&A...409.1127J}
{Janssen}, K., {V{\"o}gler}, A., \& {Kneer}, F. 2003, \aap, 409, 1127

\bibitem[{{Jefferies} {et~al.}(2006){Jefferies}, {McIntosh}, {Armstrong},
  {Bogdan}, {Cacciani}, \& {Fleck}}]{2006ApJ...648L.151J}
{Jefferies}, S.~M., {McIntosh}, S.~W., {Armstrong}, J.~D., {et~al.} 2006,
  \apjl, 648, L151

\bibitem[{{Judge}(2006)}]{2006ASPC..354..259J}
{Judge}, P. 2006, in Astronomical Society of the Pacific Conference Series,
  Vol. 354, Solar MHD Theory and Observations: A High Spatial Resolution
  Perspective, ed. J.~{Leibacher}, R.~F. {Stein}, \& H.~{Uitenbroek}, 259--+

\bibitem[{{Judge} {et~al.}(2001){Judge}, {Tarbell}, \&
  {Wilhelm}}]{2001ApJ...554..424J}
{Judge}, P.~G., {Tarbell}, T.~D., \& {Wilhelm}, K. 2001, \apj, 554, 424

\bibitem[{{Kalkofen}(1997)}]{1997ApJ...486L.145K}
{Kalkofen}, W. 1997, \apjl, 486, L145+

\bibitem[{{Langangen} {et~al.}(2007){Langangen}, {Carlsson}, {Rouppe van der
  Voort}, {Hansteen}, \& {De Pontieu}}]{langangen_07}
{Langangen}, O., {Carlsson}, M., {Rouppe van der Voort}, L., {Hansteen}, V., \&
  {De Pontieu}, B. 2007, \apj~in press (arXiv 0710.0247)

\bibitem[{{Leenaarts} {et~al.}(2006){Leenaarts}, {Rutten}, {Carlsson}, \&
  {Uitenbroek}}]{2006A&A...452L..15L}
{Leenaarts}, J., {Rutten}, R.~J., {Carlsson}, M., \& {Uitenbroek}, H. 2006,
  \aap, 452, L15

\bibitem[{{Linsky} {et~al.}(1979){Linsky}, {Hunten}, {Sowell}, {Glackin}, \&
  {Kelch}}]{1979ApJS...41..481L}
{Linsky}, J.~L., {Hunten}, D.~M., {Sowell}, R., {Glackin}, D.~L., \& {Kelch},
  W.~L. 1979, \apjs, 41, 481

\bibitem[{{Linsky} {et~al.}(1970){Linsky}, {Teske}, \&
  {Wilkinson}}]{1970SoPh...11..374L}
{Linsky}, J.~L., {Teske}, R.~G., \& {Wilkinson}, C.~W. 1970, \solphys, 11, 374

\bibitem[{{Lites}(1984)}]{1984ApJ...277..874L}
{Lites}, B.~W. 1984, \apj, 277, 874

\bibitem[{{Lites} {et~al.}(1993){Lites}, {Rutten}, \&
  {Kalkofen}}]{1993ApJ...414..345L}
{Lites}, B.~W., {Rutten}, R.~J., \& {Kalkofen}, W. 1993, \apj, 414, 345

\bibitem[{{Lyot}(1933)}]{lyot_33}
{Lyot}, B. 1933, C. R. Acad. Sci., Paris, 197, 1593

\bibitem[{{Mein}(1971)}]{1971SoPh...20....3M}
{Mein}, P. 1971, \solphys, 20, 3

\bibitem[{{Mein}(1991)}]{1991A&A...248..669M}
{Mein}, P. 1991, \aap, 248, 669

\bibitem[{{Mein}(2002)}]{2002A&A...381..271M}
{Mein}, P. 2002, \aap, 381, 271

\bibitem[{{Mein} {et~al.}(2000){Mein}, {Briand}, {Heinzel}, \&
  {Mein}}]{2000A&A...355.1146M}
{Mein}, P., {Briand}, C., {Heinzel}, P., \& {Mein}, N. 2000, \aap, 355, 1146

\bibitem[{{Noyes}(1967)}]{1967IAUS...28..293N}
{Noyes}, R.~W. 1967, in IAU Symposium, Vol.~28, Aerodynamic Phenomena in
  Stellar Atmospheres, ed. R.~N. {Thomas}, 293--+

\bibitem[{{Pasachoff} {et~al.}(1968){Pasachoff}, {Noyes}, \&
  {Beckers}}]{1968SoPh....5..131P}
{Pasachoff}, J.~M., {Noyes}, R.~W., \& {Beckers}, J.~M. 1968, \solphys, 5, 131

\bibitem[{{Pietarila} {et~al.}(2007{\natexlab{a}}){Pietarila}, {Socas-Navarro},
  \& {Bogdan}}]{2007arXiv0707.1310P}
{Pietarila}, A., {Socas-Navarro}, H., \& {Bogdan}, T. 2007{\natexlab{a}},
  \apj~in press (arXiv 0707.1310)

\bibitem[{{Pietarila} {et~al.}(2007{\natexlab{b}}){Pietarila}, {Socas-Navarro},
  \& {Bogdan}}]{2007ApJ...663.1386P}
{Pietarila}, A., {Socas-Navarro}, H., \& {Bogdan}, T. 2007{\natexlab{b}}, \apj,
  663, 1386

\bibitem[{{Pietarila} {et~al.}(2006){Pietarila}, {Socas-Navarro}, {Bogdan},
  {Carlsson}, \& {Stein}}]{2006ApJ...640.1142P}
{Pietarila}, A., {Socas-Navarro}, H., {Bogdan}, T., {Carlsson}, M., \& {Stein},
  R.~F. 2006, \apj, 640, 1142

\bibitem[{{Puschmann} {et~al.}(2006){Puschmann}, {Kneer}, {Seelemann}, \&
  {Wittmann}}]{2006A&A...451.1151P}
{Puschmann}, K.~G., {Kneer}, F., {Seelemann}, T., \& {Wittmann}, A.~D. 2006,
  \aap, 451, 1151

\bibitem[{{Qu} \& {Xu}(2002)}]{2002ChJAA...2...71Q}
{Qu}, Z.-Q. \& {Xu}, Z. 2002, Chinese Journal of Astronomy and Astrophysics, 2,
  71

\bibitem[{{Reardon} {et~al.}(2007){Reardon}, {Cauzzi}, \&
  {Rimmele}}]{2007ASPC..368..151R}
{Reardon}, K.~P., {Cauzzi}, G., \& {Rimmele}, T. 2007, in Astronomical Society
  of the Pacific Conference Series, Vol. 368, The Physics of Chromospheric
  Plasmas, ed. P.~{Heinzel}, I.~{Dorotovi{\v c}}, \& R.~J. {Rutten}, 151--+

\bibitem[{{Reardon} \& {Cavallini}(2007)}]{ibis2}
{Reardon}, K.~P. \& {Cavallini}, F. 2007, \aap~in press

\bibitem[{{Rimmele}(2004)}]{2004SPIE.5490...34R}
{Rimmele}, T.~R. 2004, in Presented at the Society of Photo-Optical
  Instrumentation Engineers (SPIE) Conference, Vol. 5490, Advancements in
  Adaptive Optics. Edited by Domenico B. Calia, Brent L. Ellerbroek, and
  Roberto Ragazzoni. Proceedings of the SPIE, Volume 5490, pp. 34-46 (2004).,
  ed. D.~{Bonaccini Calia}, B.~L. {Ellerbroek}, \& R.~{Ragazzoni}, 34--46

\bibitem[{{Rouppe van der Voort} {et~al.}(2007){Rouppe van der Voort}, {De
  Pontieu}, {Hansteen}, {Carlsson}, \& {van Noort}}]{2007ApJ...660L.169R}
{Rouppe van der Voort}, L.~H.~M., {De Pontieu}, B., {Hansteen}, V.~H.,
  {Carlsson}, M., \& {van Noort}, M. 2007, \apjl, 660, L169

\bibitem[{{Rutten}(2006)}]{2006ASPC..354..276R}
{Rutten}, R.~J. 2006, in Astronomical Society of the Pacific Conference Series,
  Vol. 354, Solar MHD Theory and Observations: A High Spatial Resolution
  Perspective, ed. J.~{Leibacher}, R.~F. {Stein}, \& H.~{Uitenbroek}, 276--+

\bibitem[{{Rutten}(2007)}]{2007ASPC..368...27R}
{Rutten}, R.~J. 2007, in Astronomical Society of the Pacific Conference Series,
  Vol. 368, The Physics of Chromospheric Plasmas, ed. P.~{Heinzel},
  I.~{Dorotovi{\v c}}, \& R.~J. {Rutten}, 27--+

\bibitem[{{Rutten} {et~al.}(2004){Rutten}, {de Wijn}, \&
  {S{\"u}tterlin}}]{2004A&A...416..333R}
{Rutten}, R.~J., {de Wijn}, A.~G., \& {S{\"u}tterlin}, P. 2004, \aap, 416, 333

\bibitem[{{Rutten} \& {Uitenbroek}(1991)}]{1991SoPh..134...15R}
{Rutten}, R.~J. \& {Uitenbroek}, H. 1991, \solphys, 134, 15

\bibitem[{{Shine} \& {Linsky}(1972)}]{1972SoPh...25..357S}
{Shine}, R.~A. \& {Linsky}, J.~L. 1972, \solphys, 25, 357

\bibitem[{{Shine} \& {Linsky}(1974)}]{1974SoPh...39...49S}
{Shine}, R.~A. \& {Linsky}, J.~L. 1974, \solphys, 39, 49

\bibitem[{{Smith} \& {Drake}(1987)}]{1987A&A...181..103S}
{Smith}, G. \& {Drake}, J.~J. 1987, \aap, 181, 103

\bibitem[{{Socas-Navarro}(2005)}]{2005ApJ...631L.167S}
{Socas-Navarro}, H. 2005, \apjl, 631, L167

\bibitem[{{Socas-Navarro} {et~al.}(2006){Socas-Navarro}, {Elmore}, {Pietarila},
  {Darnell}, {Lites}, {Tomczyk}, \& {Hegwer}}]{2006SoPh..235...55S}
{Socas-Navarro}, H., {Elmore}, D., {Pietarila}, A., {et~al.} 2006, \solphys,
  235, 55

\bibitem[{{Socas-Navarro} {et~al.}(2000){Socas-Navarro}, {Trujillo Bueno}, \&
  {Ruiz Cobo}}]{2000ApJ...530..977S}
{Socas-Navarro}, H., {Trujillo Bueno}, J., \& {Ruiz Cobo}, B. 2000, \apj, 530,
  977

\bibitem[{{Socas-Navarro} \& {Uitenbroek}(2004)}]{2004ApJ...603L.129S}
{Socas-Navarro}, H. \& {Uitenbroek}, H. 2004, \apjl, 603, L129

\bibitem[{{Stenflo}(1994)}]{1994ASSL..189.....S}
{Stenflo}, J.~O., ed. 1994, Astrophysics and Space Science Library, Vol. 189,
  {Solar magnetic fields: polarized radiation diagnostics}

\bibitem[{{Title}(1966)}]{title_66}
{Title}, A.~M. 1966, Mount Wilson and Palomar Observatories, Pasadena

\bibitem[{{Tritschler} {et~al.}(2002){Tritschler}, {Schmidt}, {Langhans}, \&
  {Kentischer}}]{2002SoPh..211...17T}
{Tritschler}, A., {Schmidt}, W., {Langhans}, K., \& {Kentischer}, T. 2002,
  \solphys, 211, 17

\bibitem[{{Tziotziou} {et~al.}(2006){Tziotziou}, {Tsiropoula}, {Mein}, \&
  {Mein}}]{2006A&A...456..689T}
{Tziotziou}, K., {Tsiropoula}, G., {Mein}, N., \& {Mein}, P. 2006, \aap, 456,
  689

\bibitem[{{Uitenbroek}(1989)}]{1989A&A...213..360U}
{Uitenbroek}, H. 1989, \aap, 213, 360

\bibitem[{{Uitenbroek}(2006{\natexlab{a}})}]{2006ASPC..354..313U}
{Uitenbroek}, H. 2006{\natexlab{a}}, in Astronomical Society of the Pacific
  Conference Series, Vol. 354, Solar MHD Theory and Observations: A High
  Spatial Resolution Perspective, ed. J.~{Leibacher}, R.~F. {Stein}, \&
  H.~{Uitenbroek}, 313--+

\bibitem[{{Uitenbroek}(2006{\natexlab{b}})}]{2006ApJ...639..516U}
{Uitenbroek}, H. 2006{\natexlab{b}}, \apj, 639, 516

\bibitem[{{Uitenbroek} {et~al.}(2006){Uitenbroek}, {Balasubramaniam}, \&
  {Tritschler}}]{2006ApJ...645..776U}
{Uitenbroek}, H., {Balasubramaniam}, K.~S., \& {Tritschler}, A. 2006, \apj,
  645, 776

\bibitem[{{van Noort} {et~al.}(2005){van Noort}, {Rouppe van der Voort}, \&
  {L{\"o}fdahl}}]{2005SoPh..228..191V}
{van Noort}, M., {Rouppe van der Voort}, L., \& {L{\"o}fdahl}, M.~G. 2005,
  \solphys, 228, 191

\bibitem[{{van Noort} \& {Rouppe van der Voort}(2006)}]{2006ApJ...648L..67V}
{van Noort}, M.~J. \& {Rouppe van der Voort}, L.~H.~M. 2006, \apjl, 648, L67

\bibitem[{{Vecchio} {et~al.}(2007{\natexlab{a}}){Vecchio}, {Cauzzi}, \&
  {Reardon}}]{vecchio_shocks}
{Vecchio}, A., {Cauzzi}, G., \& {Reardon}, K.~P. 2007{\natexlab{a}}, \aap, to
  be submitted

\bibitem[{{Vecchio} {et~al.}(2007{\natexlab{b}}){Vecchio}, {Cauzzi}, {Reardon},
  {Janssen}, \& {Rimmele}}]{2007A&A...461L...1V}
{Vecchio}, A., {Cauzzi}, G., {Reardon}, K.~P., {Janssen}, K., \& {Rimmele}, T.
  2007{\natexlab{b}}, \aap, 461, L1

\bibitem[{{Wedemeyer-B{\"o}hm} {et~al.}(2007){Wedemeyer-B{\"o}hm}, {Steiner},
  {Bruls}, \& {Rammacher}}]{2007ASPC..368...93W}
{Wedemeyer-B{\"o}hm}, S., {Steiner}, O., {Bruls}, J., \& {Rammacher}, W. 2007,
  in Astronomical Society of the Pacific Conference Series, Vol. 368, The
  Physics of Chromospheric Plasmas, ed. P.~{Heinzel}, I.~{Dorotovi{\v c}}, \&
  R.~J. {Rutten}, 93--+

\bibitem[{{W{\"o}ger}(2006)}]{Woeger_thesis}
{W{\"o}ger}, F. 2006, PhD thesis, Albert-Ludwigs Universit{\"a}t Freiburg,
  Freiburg, Germany

\bibitem[{{W{\"o}ger} {et~al.}(2006){W{\"o}ger}, {Wedemeyer-B{\"o}hm},
  {Schmidt}, \& {von der L{\"u}he}}]{2006A&A...459L...9W}
{W{\"o}ger}, F., {Wedemeyer-B{\"o}hm}, S., {Schmidt}, W., \& {von der
  L{\"u}he}, O. 2006, \aap, 459, L9

\end{thebibliography}
\end{document}